\documentclass[aip,rsi,groupedaddress,reprint
]{revtex4-1}

\usepackage{soul}
\usepackage{rotating}
\usepackage{siunitx} 
\usepackage{color}
\usepackage[usenames,dvipsnames]{xcolor}
\usepackage{mathtools}
\usepackage{mathrsfs}
\usepackage{amsmath}
\usepackage{amsthm} 
\usepackage{amssymb}	
\usepackage{epsfig}
\usepackage{array}
\usepackage{float}
\usepackage{wrapfig}
\usepackage{graphicx}
\usepackage{epstopdf}
\usepackage{dcolumn}
\usepackage{enumerate}
\usepackage[shortlabels]{enumitem}
            \setlist[enumerate, 1]{1)}
\usepackage{enumitem}
\usepackage{textcomp}
\usepackage{appendix}
\usepackage[colorlinks=true,linkcolor=blue,citecolor=blue,urlcolor=blue,bookmarksopen]{hyperref} 
\usepackage{tabularx,ragged2e,booktabs}
\newcolumntype{Y}{>{\RaggedRight\arraybackslash}X}
\usepackage{multirow}

\DeclareFontEncoding{LS1}{}{}
\DeclareFontSubstitution{LS1}{stix}{m}{n}

\let\emptyset\varnothing

\newcommand{\gv}[1]{\ensuremath{\mbox{\boldmath$ #1 $}}}
\newcommand{\abs}[1]{\left| #1 \right|} 
\newcommand{\grad}[1]{\gv{\nabla} #1} 

\newcommand{\green}{\color{blue!10!black!60!green}}
\newcommand{\magenta}{\color{magenta}}

\newcommand{\ket}[1]{\left| #1 \right>} 

\newcommand{\plane}[1]{$#1$-plane} 

\newcommand{\zhat}{\hat{z}}

\newcommand{\Avec}{\vec{\boldsymbol{A}}}
\newcommand{\Bvec}{\vec{\boldsymbol{B}}}
\newcommand{\Evec}{\vec{\boldsymbol{E}}}

\newcommand{\rvec}{\vec{\boldsymbol{r}}}

\newcommand{\qch}{q} 
 
\newcommand{\IonBep}{$^{9}$\textrm{Be}$^{+}$}

\newcommand{\tptt}{$2p\,^2P_{3/2}$}
\newcommand{\tsot}{{$2s\,^2S_{1/2}$}}
\newcommand{\osz}{$^1S_{0}\,$}
\newcommand{\opo}{$^1P_1\,$}
\newcommand{\HH}{\mathcal{H}}

\newcommand{\eg}{e.g.}

\newcommand{\etal}{et al.}

\begin{document}




\title{Site-resolved imaging of beryllium ion crystals in a high-optical-access Penning trap with inbore optomechanics}

\author{H. Ball}
\altaffiliation[Present Address: ]{Q-CTRL Pty Ltd, Sydney, NSW 2000, Australia}
\thanks{These three authors contributed equally to this work}
\affiliation{ARC Centre for Engineered Quantum Systems, School of Physics, The University of Sydney, NSW 2006, Australia}

\author{Ch. D. Marciniak}
\thanks{These three authors contributed equally to this work}
\affiliation{ARC Centre for Engineered Quantum Systems, School of Physics, The University of Sydney, NSW 2006, Australia}

\author{R. N. Wolf}
\thanks{These three authors contributed equally to this work}
\affiliation{ARC Centre for Engineered Quantum Systems, School of Physics, The University of Sydney, NSW 2006, Australia}

\author{A. T.-H. Hung}
\altaffiliation[Present Address:]{ Department of Electrical and Computer Engineering, The University of California, Los Angeles, CA, USA}
\affiliation{ARC Centre for Engineered Quantum Systems, School of Physics, The University of Sydney, NSW 2006, Australia}

\author{K. Pyka}
\altaffiliation[Present Address:]{ Hella Corporation, Berlin, Germany}
\affiliation{ARC Centre for Engineered Quantum Systems, School of Physics, The University of Sydney, NSW 2006, Australia}

\author{M. J. Biercuk}
\email[Corresponding author: ]{michael.biercuk@sydney.edu.au }
\affiliation{ARC Centre for Engineered Quantum Systems, School of Physics, The University of Sydney, NSW 2006, Australia}



\begin{abstract}
We present the design, construction and characterization of an experimental system capable of supporting a broad class of quantum simulation experiments with hundreds of spin qubits using \IonBep~ions in a Penning trap. This article provides a detailed overview of the core optical and trapping subsystems, and their integration. We begin with a description of a dual-trap design separating loading and experimental zones and associated vacuum infrastructure design. The experimental-zone trap electrodes are designed for wide-angle optical access (e.g. for lasers  used to engineer spin-motional coupling across large ion crystals) while simultaneously providing a harmonic trapping potential. We describe a near-zero-loss liquid-cryogen-based superconducting magnet, employed in both trapping and establishing a quantization field for ion spin-states, and equipped with a dual-stage remote-motor LN$_2$/LHe recondenser. Experimental measurements using a nuclear magnetic resonance (NMR) probe demonstrate  part-per-million homogeneity over \SI{7}{\milli\meter}-diameter cylindrical volume, with no discernible effect on the measured NMR linewidth from pulse-tube operation.   Next we describe a custom-engineered inbore optomechanical system which delivers ultraviolet (UV) laser light to the trap, and supports multiple aligned optical objectives for top- and sideview imaging in the experimental trap region.  We describe design choices including the use of non-magnetic goniometers and translation stages for precision alignment.  Further, the optomechanical system integrates UV-compatible fiber optics which decouple the system's alignment from remote light sources.  Using this system we present site-resolved images of ion crystals and demonstrate the ability to realize both planar and three-dimensional ion arrays via control of rotating wall electrodes and radial laser beams. Looking to future work, we include interferometric vibration measurements demonstrating root-mean-square trap motion of $\sim \SI{33}{\nano\meter}$ ($\sim \SI{117}{\nano\meter}$) in the axial (transverse) direction; both values can be reduced when operating the magnet in free-running mode.  The paper concludes with an outlook towards extensions of the experimental setup, areas for improvement, and future experimental studies. 
\end{abstract}

\maketitle



\section{Introduction}
\label{Sec:Introduction:Simulation}

Over recent decades significant progress, both theoretical and experimental, has been made towards building a universal quantum computer. Realizing such a device, however, remains a challenge requiring breakthroughs in many areas of science and engineering \cite{Ladd2010}. Fortunately, this is not the sole approach for the implementation of important computational tasks where quantum advantages may be gained.  Here we focus on the concept of an analog quantum simulator~\cite{Feynman1982}; namely, a controllable quantum system that \emph{mimics} the behaviour, or evolution, of another less accessible system of interest. The simulator system must possess a Hamiltonian that captures the important features of the system being studied, and must be controlled, manipulated and measured in a sufficiently precise manner. Although lacking the universality of a general-purpose quantum computer, such problem-specific ``analog'' machines may provide computational advantages at intermediate scales that may be easier to construct than comparable universal machines.  It is therefore expected that practical quantum simulation may become a reality well before fully-fledged quantum computers~\cite{GeorgescuRevModPhys2014}. 

The potential utility of mesoscale quantum simulators has grown as the technologies required for coherent manipulation of quantum systems have matured, and early practical applications have become feasible~\cite{Ladd2010, BulutaRepProgPhys2011}.  A substantial body of proof-of-principle experiments has already been realized using a variety of physical systems, including nuclear spins~\cite{KassalAnnuRevPhysChem2011, LuPhysChemChemPhys2012}, photonic systems~\cite{Lanyon2010, WaltherNatPhys2012,Matthews2013}, superconducting circuits~\cite{NeeleyScience2009, HouckNatPhys2012,Marcos2013}, cold atoms~\cite{GreinerNature2002, JakschAnnals2005, LewensteinAdvPhys2007, BlochRevModPhys2008,Simon2011, Struck2011}, and trapped ions~\cite{LeibfriedPRL2002, FriedenauerNatPhys2008, GerritsmaNature2010, MonroeNatPhys2010, Lanyon2011, BlattNatPhys2012, SchneiderRepProgPhys2012, Britton2012}. New experimental demonstrations continue to validate the underlying potential of this computational paradigm for the simulation of physical systems.

One class of device attracting interest comprises systems which support controllable Ising interactions exhibiting magnetic frustration~\cite{ElliotPRL1970}.  Here competing interactions between spins on {\em e.g.} a  triangular lattice with antiferromagnetic couplings prevent the realization of configurations which simultaneously minimize the energies of all pairwise interactions.  In the classical limit this behaviour is posited to be central to understanding many complex systems, from social~\cite{WassermanBook1994} and neural networks~\cite{DorogovtsevRevModPhys2008} to protein folding~\cite{BryngelsonProcNatl1987} and magnetism~\cite{DiepWorldSci2005,MoessnerPhys.Today2006}. 

Quantum superposition and entanglement between spins in such systems give rise to long-range quantum spin correlations believed to underlie many exotic phenomena, such as quantum phase transitions, many-body localization, and potentially even high-temperature superconductivity \cite{Moessner2000, Balents2011, banerjee2016proximate, AndersonScience2011, MoessnerPhys.Today2006,MoessnerRevModPhys2003}. In quantum networks, frustration leads to massively entangled and highly degenerate ground states with excess entropy even at zero temperature, underpinning exotic materials such as quantum spin liquids and spin glasses~\cite{BinderRevModPhys1986,SachdevBook1999,DawsonPhysRevA2004,NormandPhysRevB2008,MonroeNatPhys2010, NandkishoreAnnuRev2015,BelitzRevModPhys2005}.

Of the experimental platforms referenced above, trapped ions have several advantages in realizing the physics of Ising quantum spin simulations, such as high interconnectivity and high fidelity spin control. Many experimental demonstrations have already been conducted using radio-frequency Paul traps~\cite{FriedenauerNatPhys2008,MonroeNatPhys2010,Islam03052013,Jurcevic2014,Zhang2017}. More recently, new capabilities for the engineering of spin interactions in Penning traps have emerged, opening novel capabilities at the scale of hundreds of interacting spins~\cite{Britton2012, Sawyer2012, BohnetScience2016, GarttnerNatPhys2017}, at the cost of controls limited to global interactions. This hardware limitation can be at least partially offset by using quantum control techniques~\cite{HayesNJP2014, korenblit2012}, extending the reach of  programmable quantum simulation.

Penning traps employ a combination of static electric and magnetic fields to confine charged particles. This provides advantages for trapping large, stable ion crystals for quantum simulation experiments~\cite{Porras2004,Porras2006}. Under appropriate experimental conditions, laser-cooled ions in Penning traps self-assemble into two-dimensional triangular lattices~\cite{Mitchell1998}, and are amenable to high-fidelity spin-state control~\cite{BiercukQIC2009}, long trapping times, and straightforward mechanisms for generating transverse-field Ising interactions\cite{FreericksPRA2013,FreericksPRA2015}. These capabilities have been demonstrated in a number of experiments, including engineering and benchmarking of long-range Ising interactions~\cite{Britton2012, Sawyer2012}, observation of quantum entanglement between spin-squeezed states~\cite{BohnetScience2016}, and measurement of out-of-time-order correlations as a signature of many-body quantum correlations~\cite{GarttnerNatPhys2017}.

In this work we present the design and construction of an experimental setup that addresses key technical challenges in building a mesoscale quantum simulator in a Penning trap with hundreds of qubits.  Our work builds on the pioneering efforts of the Bollinger group at NIST~\cite{Brewer1988,
Bollinger1991,
Heinzen1991,
Bollinger1993,
Huang1998,
Mitchell1998,
Mitchell2001,
Kriesel2002,
Bollinger2003,
Jensen2004,
Hasegawa2005,
BiercukQIC2009,
Biercuk2010,
Shiga2011,
Britton2012,
Sawyer2012,
Sawyer2014,
Sawyer2015,
BrittonMagnet2016,
BohnetScience2016,
Torrisi2016,
ShankarPRA2017,
GarttnerNatPhys2017},
 and the Thompson group at Imperial College London~\cite{ThompsonAppPhysB2012,
ThompsonAppPhysB2014,
Goodwin2016,
mavadia2013} over the last two decades. We describe our system design in detail and provide demonstrations of system functionality, realizing site-resolved images of 2D and 3D crystals of beryllium ions. Our presentation includes discussion of design drivers, technical approaches, and characterization of subsystem performance.
We describe in detail three subsystems in our setup: 
(i) the horizontally oriented high-optical-access ion trap and vacuum chamber, (ii) a near-zero-loss wide-bore magnet with dual-stage reliquefier, and (iii) non-magnetic, inbore optomechanics for initial characterization of the system in performing foundational tasks outlined above (with the exception of Raman beam delivery). Overall we aim to provide a comprehensive account of the relevant system features, highlighting unique aspects where they arise, noting that many of the hardware design features of existing systems have remained unpublished. 

The remainder of this paper is organized as follows. Section~\ref{Sec:Berylliumtransitions} summarizes the
atomic structure of beryllium to identify the relevant optical and millimeter-wave transitions.
Section~\ref{Sec:ExperimentalDesign} describes the experimental setup in detail, addressing the three major subsystems identified above.  Section~\ref{Sec:Results} presents details on trap operation, data on the production, confinement and imaging of cold ions,  analysis of their crystal structures, and interferometric measurements of the mechanical stability of the trap and associated optomechanics. Section~\ref{Sec:Conclusion} concludes with a brief summary and a detailed discussion of future work, including plans for Raman beam integration, active wavefront alignment, sub-Doppler cooling, and incorporation of precision trap positioning capabilities.



\section{Light-matter interaction in Beryllium-9} \label{Sec:Berylliumtransitions}

Quantum simulation experiments with ion crystals typically leverage a variety of light-matter interactions for tasks including:  (i) Resonantly enhanced multi-photon photoionization; (ii) Doppler cooling; (iii) repumping ions that leave the Doppler-cooling cycling transition; (iv) Raman interactions to mediate spin-spin couplings; and lastly (v) millimeter-wave interactions to drive qubit transitions. In this section we identify relevant aspects of the beryllium energy level structures that motivate downstream design choices.  This summary is supplemented by detailed atomic physics calculations presented in Appendix~\ref{Appendix:BeLevels}.

\begin{figure}[b]
\begin{center}
\includegraphics[width=8.5cm]{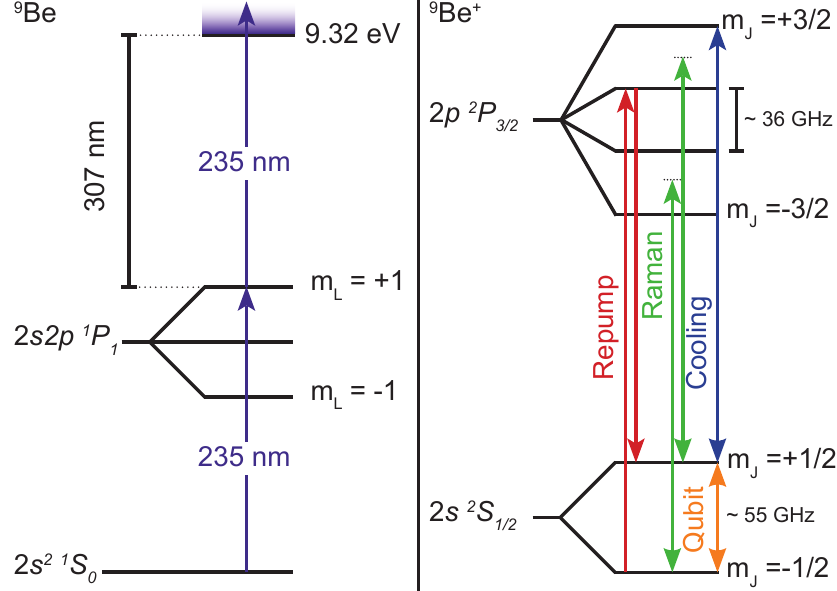}
\caption{Energy levels and relevant transitions in atomic and ionized beryllium in a $\SI{2}{\tesla}$ magnetic field (not drawn to scale). Left: Resonantly enhanced, two-photon ionization process of one valence electron from $^1$S$_0$ to $^1$P$_1$ in neutral $^9$Be, the excited-state-to-continuum wavelength is $\approx\SI{307}{\nano\meter}$. Right: \IonBep\ energy level diagram with transitions for Doppler cooling, repump, qubit, and Raman transitions shown. Only the $m_I=+3/2$ levels are shown. This diagram is only valid after the nuclear dynamics have been frozen out by optically pumping the $m_I$ states to the $+3/2$ manifold through the cooling and repump transitions with properly chosen polarization. }
\label{fig:EnergyLevels}
\end{center}
\end{figure}

The production of ions from neutral beryllium constitutes the first step of any experiment and may proceed through various approaches. We focus on photoionization as it provides elemental selectivity, efficient on-axis ionization, and can reduce the possibility of electrode and insulator charging or the production of trapped contaminant ions from background gas as may occur when using electron impact ionization. In addition the position of the laser beam can be easily controlled ensuring atoms are ionized close to the trap center where they are more efficiently laser-cooled. 

Beryllium-9 has a nuclear spin of $I=3/2$, which couples to the electron spin of $S=1/2$ to form hyperfine structure. The relevant energy level diagram for neutral $^9$Be is shown on the left side of Fig.~\ref{fig:EnergyLevels} including magnetic field corrections from an approximately $\SI{2}{\tesla}$ superconducting magnet employed to provide radial confinement in the Penning trap, see Appendix~\ref{Sec:IdealPenningTrap} for a description of the physics of ion confinement in a Penning trap. The detailed transition frequencies required for photoionization of neutral beryllium have been calculated here using standard atomic physics techniques and are summarized in Table~\ref{Table:BeCalcs} in Appendix~\ref{Appendix:BeLevels}.

Direct photoionization of $^9$Be requires light with a wavelength of \SI{133}{\nano\meter}, which lies in the inconvenient vacuum ultraviolet spectral region. It is therefore advantageous to utilize a resonantly enhanced two-photon photoionization scheme where the efficiency of a multi-photon process is augmented by tuning the first excitation to a resonant transition of $^9$Be before exciting to the continuum\cite{Lucas2004}. We generate the required \SI{235}{\nano\meter} light by doubling the output of a high-power diode laser with emission at \SI{470}{\nano\meter}.  The details of this laser setup will be published elsewhere.

The relevant electronic transitions in \IonBep\ for qubit manipulation, Doppler cooling, repumping, and Raman interactions are illustrated on the right hand side of Fig.~\ref{fig:EnergyLevels}. For simplicity the electronic energy levels are depicted in the $\ket{m_I=+3/2,m_J}$ basis, omitting hyperfine states that do not take part in the transitions we consider. The qubit is encoded in the Zeeman-split ground states of the valence electron spin of \IonBep, namely the \tsot\ states parallel $\ket{\uparrow}=\ket{m_J=+1/2}$ and anti-parallel $\ket{\downarrow}=\ket{m_J=-1/2}$ to the confining magnetic field of the Penning trap. The $B_0\approx\SI{2}{\tesla}$ magnetic field used in this setup produces an energy splitting of $\approx\SI{55}{\giga\hertz}$. This transition is addressed using a custom-designed U-band horn antenna and elliptical mirror which will be described elsewhere.

The primary optical transitions in \IonBep\ are all around \SI{313}{\nano\meter}, including the Doppler cooling transition. In addition to cooling, the Doppler transition is used for state-selective readout, exploiting the fact that the $\ket{\uparrow}$ state will scatter \SI{313}{\nano\meter} photons, while the $\ket{\downarrow}$ is dark.
A repump transition is driven from $\ket{\downarrow}$ to $\ket{^2P_{3/2}\,m_J=+1/2}$, which optically pumps all ions to the $\ket{\uparrow}$ state during qubit-state initialization. The final optical interaction is used to engineer Ising-type Hamiltonians by generating effective spin-spin interactions between ions across the crystal.  This is accomplished using Raman lasers which off-resonantly excite modes of motion in the trapped ions~\cite{Britton2012} via an optical dipole force. The detailed frequencies relevant to our experiment appear in Table~\ref{Table:BeCalcs}. The laser system used to generate the various laser beams broadly follows the approach demonstrated by Wilson \etal~\cite{Wilson2011}.



\section{Experimental System Design}
\label{Sec:ExperimentalDesign} 

\begin{figure*}
		\includegraphics[width=\textwidth]{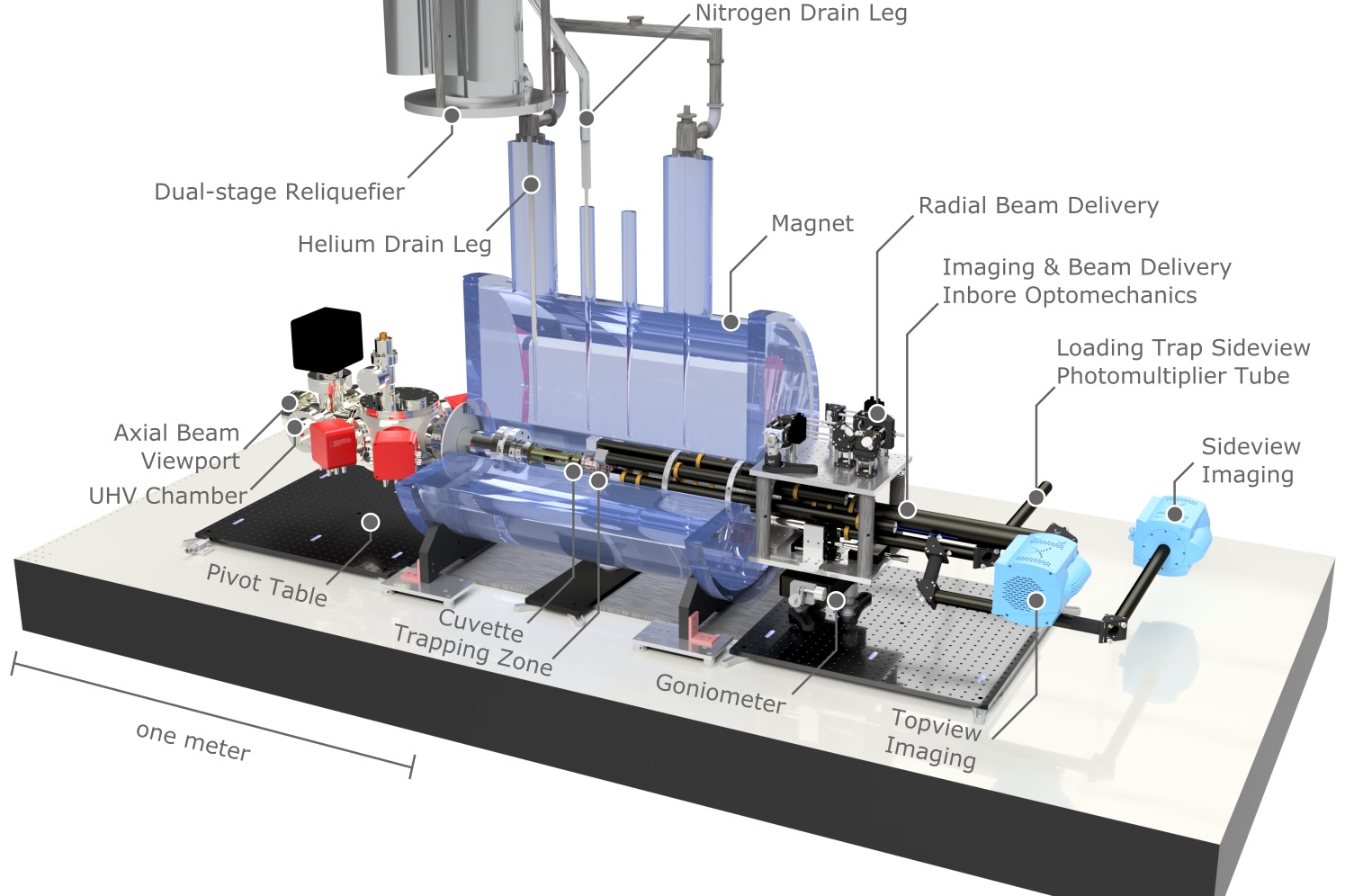}
	 \caption{Simplified experimental setup overview. Core subsystems include the ultra-high-vacuum (UHV) chamber and trap (left), magnet and reliquefier (center), and optomechanics and imaging system (right). Axial beam delivery and vacuum chamber mounting are omitted for clarity. The entire assembly is constructed on a $\SI{1.5}{\meter}\times\SI{3}{\meter}$ optical table. Laser sources are constructed on a separate optical table  (not shown).}
	\label{Fig:SetupOverview}
\end{figure*}

This section details the design drivers and, consequently, choices made in realizing our experimental platform, see Fig.~\ref{Fig:SetupOverview}. This includes constraints imposed by the nature of the light-matter interactions outlined above, as well as by the underlying trapping mechanisms for the Penning trap itself.  Our objective is to provide a detailed narrative description of the design process and the solutions employed as we are not aware of such information being published for systems from other experimental teams.

To begin we describe optical access requirements for delivering laser beams into the trap. Typically, both repump and axial Doppler beams will be oriented parallel to the trap axis and magnetic field, with an additional transverse Doppler beam incorporated for intensity-gradient cooling \cite{Itano1982,Torrisi2016}. This configuration minimizes coupling between axial and in-plane degrees of freedom. Consequently, we require both on-axis, and transverse optical access through the trap electrodes and vacuum chamber. Next, Raman beams will enter at a shallow angle symmetrically about the crystal plane with a full opening angle of $\theta_R$, to produce a moving standing wave propagating along the trap axis. To drive the qubit transition with efficient magnetic-dipole coupling, millimeter waves must reach the trap center with magnetic field orientation transverse to the quantization axis. This may be achieved using either transverse or axial access, though we opt for the former in order to maximize access for axial imaging. Finally, apertures for collecting ion fluorescence are necessary for crystal analysis and state detection. Topview (on-axis) imaging can in principle provide spin-state determination for each ion in a single-plane crystal, while sideview (transverse) imaging enables unambiguous determination of the crystal conformation. 

Achieving the nominal optical access described above, however, faces the following challenges: First, introducing irregularities such as apertures into the trap electrode structure distorts the harmonicity of the trapping potential and can also impact the homogeneity of the magnetic field. Next, ingress and egress pathways for light are tightly constrained by the inner diameter of the bore of the superconducting solenoid magnet, whose use is common in precision metrology applications requiring high field homogeneity and stability. In this section we describe the design and assembly of a system meeting these competing demands. We focus on three primary subsystems: the Penning trap structure and associated ultra-high-vacuum chamber, the near-zero-loss superconducting magnet, and the inbore optomechanics for laser delivery and ion imaging.  Throughout we will use language derived from international standards for machining when describing tolerances, mating fits, and the like.

%
\subsection{Penning trap system}
\label{Sec:PenningTrapSystem}

\begin{figure}[t]
 \centering
 \begin{tabular}{c}
 \includegraphics[width=8.5cm]{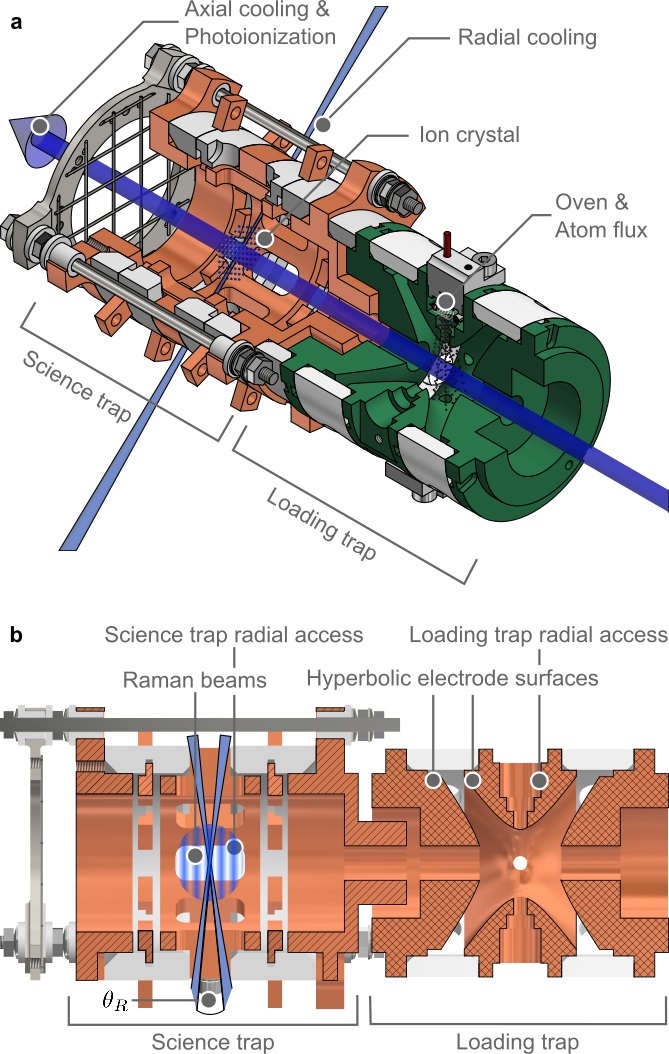}\\
 \end{tabular}
\caption{Trap assembly, and laser orientation overview. (a) Model images of trap electrodes (colored) and Macor spacers (white), showing relative orientations of beryllium ovens, trapping region, and laser trajectories.  Electrodes for the science and loading traps are shown in distinct colors for clarity. (b) Cross-sectional view of inner electrode surface geometry (hatched) showing optical access.  All electrodes in the loading trap exhibit hyperbolic surfaces to approximate the ideal potential described in Eq.~\ref{Eq:IdealHarmonicTrappingPotential}.  The science trap electrodes are cylindrical with radial optical access ports for Raman lasers and millimeter waves highlighted.  At the far left is the wire mesh serving to close the trap electrostatically. Raman beam trajectories and optical-dipole-force-inducing wavefronts from interference under angle $\theta_R$ indicated schematically; here the beams would enter through the radial access port but are rotated into view for presentation.} 
\label{Fig:TrapAssemblyAndBeamlines}
\end{figure}

We have designed our trap as two axially stacked Penning traps, referred to as the loading trap and science trap. Ions are initially created and confined in the loading trap, then shuttled into the science trap where ions are crystallized, manipulated, and imaged. Separating these functions permits greater optical access in the science trap by removing the need for an aperature accomodating  the beryllium sources, and prevents the experimental zone from being contaminated by beryllium plating generated during ion creation.  All trap electrodes are machined from oxygen-free high-conductivity copper and assembled in a stacked formation, separated by machined Macor rings for electrical isolation.  An overview of the trap assembly is shown in Fig.~\ref{Fig:TrapAssemblyAndBeamlines}, including beryllium ovens and lasers. The operating principle of the Penning trap is summarized in Appendix~\ref{Sec:IdealPenningTrap}, and more detailed discussions of this topic can be found in Refs. \cite{Metcalf2002,Knoop2016,Vogel2018}. 

%
\subsubsection{Loading trap}
\label{Sec:LoadingTrap}

\begin{figure}[tp]
	\centering{
	\includegraphics[width=8.5cm]{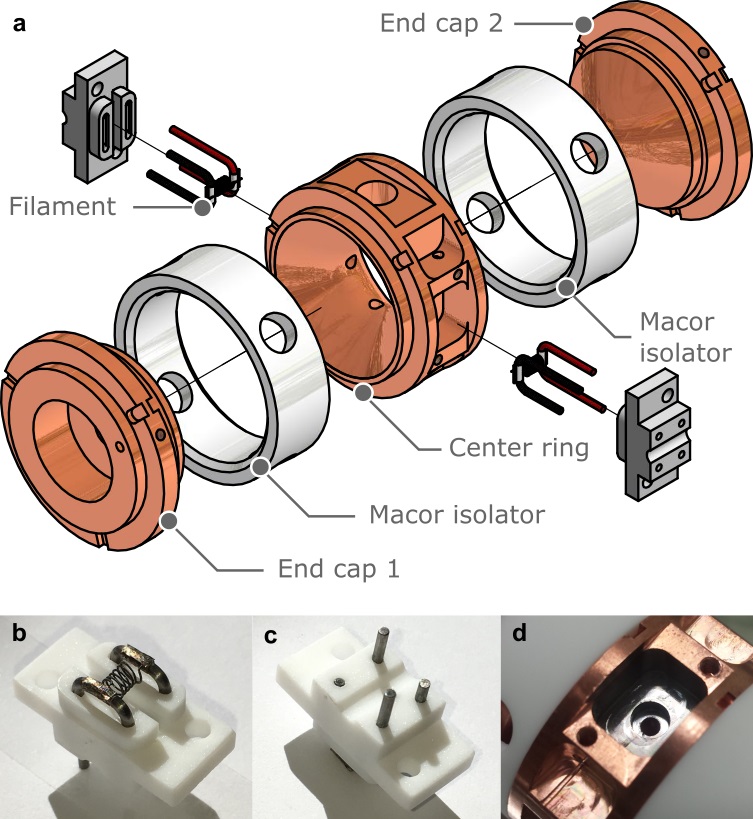}}
	\caption{(a) Loading trap design, and assembly showing relevant electrodes (colored), Macor spacers (white), and beryllium oven assemblies. (b-d) Beryllium oven assembly and mating to center ring electrode. (b) Filament coil spot-welded to tungsten electrodes glued into Macor base. Output atomic flux is roughly collimated by passing through an $\emptyset$\SI{2.0}{\milli\meter} aperture of \SI{2.3}{\milli\meter} length limiting the solid angle of the atomic flux cone into the trap center. Filament length and coil number are set to ensure that upon heating thermal expansion does not cause inadvertent shorts between adjacent coils.  Filament is spot-welded to a pair of $\emptyset$\SI{1.0}{\milli\meter} tungsten electrodes that are then glued into a Macor base using an ultra-high-vacuum-compatible ceramic adhesive. (c) Underside of assembly with tungsten electrodes protruding through Macor base. Different electrode lengths identify correct polarity for heater current. (d) Cavity milled into side of the loading trap center ring for housing beryllium ovens. Beryllium plating is observed on interior surfaces of oven housing after running oven.}
	\label{fig:LoadingTrap_AndOvens}
\end{figure}

Ionized beryllium is first produced and confined from a flux of neutral atoms inside the loading trap. The loading trap consists of an axisymmetric center ring and two end cap electrodes with hyperbolic surfaces, as shown in Fig.~\ref{fig:LoadingTrap_AndOvens}(a).  The central diameter of the center ring electrode is $r_0 = \SI{4.95}{\milli\meter}$ and the half-distance between the end caps is $z_0 = \SI{5.49}{\milli\meter}$.  This geometry is close to an ideal Penning trap~\cite{Major} 
comprising mathematically perfect hyperbolic electrodes, and requires only limited optical access. On-axis holes of \SI{5}{\milli\meter} diameter in both end cap electrodes serve to deliver the axial Doppler, repump, and photoionization beam into both sections of the trap, and to enable transfer of the ions between the traps. 

Two beryllium ovens serving as sources of $^9$Be are mounted in cavities milled in the exterior of the center ring of the loading trap, see Fig.~\ref{fig:LoadingTrap_AndOvens}. The ovens consist of a $\emptyset$\SI{50}{\micro\meter} beryllium wire tightly wound around a coiled filament of $\emptyset$\SI{100}{\micro\meter} tungsten.  The torque on the coil resulting from the presence of the magnetic field is minimized by choosing the alignment of the coil and the polarity of the heater current such that its magnetic moment is parallel to that of the superconducting magnet to avoid a deformation of the filament structure. For full details on the oven construction method see Ref.~\cite{HBall2018PhD}.

%

\subsubsection{Science trap}
\label{Sec:ScienceTrap}

The structure of the science trap is developed with the primary objectives of providing high optical access for laser beam delivery and ion imaging while maintaining high harmonicity of the trapping potential. Lasers must enter and exit the trapping region without clipping electrode surfaces to avoid unwanted charging effects and laser scatter. Apertures must therefore provide sufficient clearance for beam alignment, including accommodation of a large angular separation $\theta_R$ in the case of the Raman beams. The \SI{55}{\giga\hertz} or $\approx\SI{5.5}{\milli\meter}$ wavelength radiation used for qubit manipulation also sets a lower bound on its entry aperture size in order to minimize diffraction, as does the requirement of achieving high-numerical aperture imaging ($f\#\lesssim 5$).

 \begin{figure}[t]
 \includegraphics[width=8.5cm]{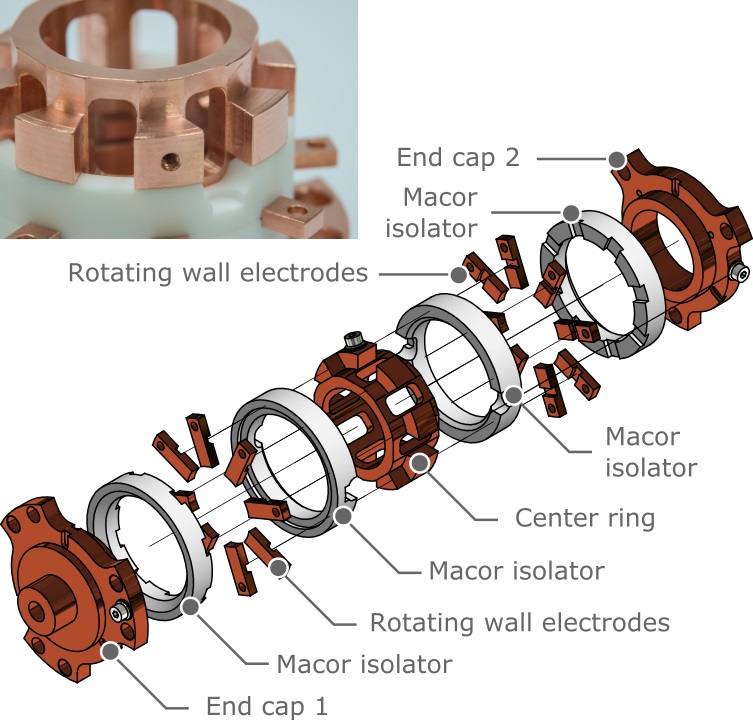}
\caption{Exploded view of the Science trap. As above metal electrodes are represented in color while Macor spacers are white. The center ring electrode has a length of \SI{14.5}{\milli\meter} and is separated from the rotating wall electrodes by \SI{1.1}{\milli\meter}. Each rotating wall electrode has a length and width of \SI{2.1}{\milli\meter} and \SI{3.9}{\milli\meter}, respectively. Their inner surface has a radius of \SI{10}{\milli\meter}, identical to the other science trap electrodes. The distance between the rotating wall electrodes and adjacent end caps is \SI{0.8}{\milli\meter}, the length of the \SI{20}{\milli\meter}-diameter part of the end caps is \SI{8.2}{\milli\meter}. After that the inner diameter of end cap 1 is reduced to \SI{5}{\milli\meter} to form a transition to the loading trap.  Inset: Photograph of science trap center ring electrode oriented to highlight eight large apertures for wide optical and millimeter wave access.} 
\label{fig:ScienceTrapCombined}
\end{figure}

\begin{figure*}
\includegraphics[width=17cm]{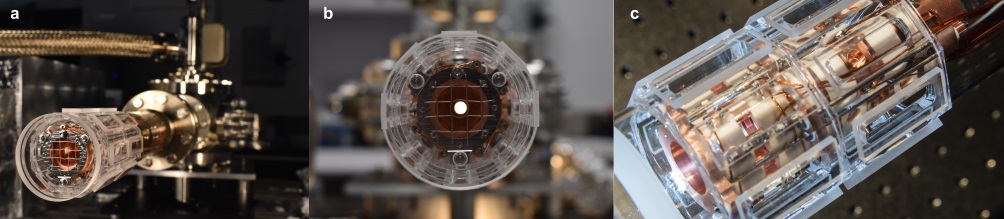}\\
\caption{Trap assembly in glass cuvette showing relationship between electrode structures, optically contacted windows, and vacuum chamber. (a) Cuvette mounted on vacuum system during vacuum bake, with pumping line still attached (top). (b) On-axis topview highlighting laser beam path on axis (bright spot) using back illumination, shielding mesh preventing cuvette charging from distorting trap potential, and eight-fold symmetric cuvette window pattern.  (c) Angled side view showing thickness of optical flat vacuum window constituting first element of topview imaging system (lower left) and alignment of windows to trap apertures.} 
\label{Fig:vacuum_assembly_with_cuvette}
\end{figure*}


The science trap design is shown in Fig.~\ref{fig:ScienceTrapCombined}. All electrode dimensions have been designed via numerical optimization to ensure the competing objectives identified above are met.  Details on the optimization procedure and resulting geometric constants appear in Appendix~\ref{Sec:IdealPenningTrap}.  

An eight-fold array of radial apertures of dimension \SI[product-units = brackets-power]{5 x 10}{\milli\metre} is machined in the center ring. This permits a maximum angular separation of $\theta_R = \SI{32}{\degree}$ for the Raman beams, $\sim1.5\times$ greater than that reported in recent work~\cite{BohnetScience2016}, and to our knowledge exceeding the range reported in any other published system.

Our design also incorporates rotating wall electrodes in an eight-fold-symmetric radial array sandwiched between the Macor rings on either side of the center ring (see Fig.~\ref{fig:ScienceTrapCombined}). These electrodes can be used to produce an azimuthally-rotating quadrupole field for locking the ion crystal rotation frequency~\cite{Huang1998}. The sixteen rotating wall electrodes are electrically connected in four groups of four, each comprising vertically- and radially-opposing electrodes.

A wire mesh of \SI{5.6}{\milli\meter} grid spacing is strung on a titanium ring structure, which is mounted on Macor spacers above end cap 2 of the science trap.  This structure closes the Penning trap electrostatically, preventing distortions of the trapping potential due to static charges on the interior surface of the enclosing glass vacuum cuvette (see Fig.~\ref{Fig:vacuum_assembly_with_cuvette}). The wire mesh is centered on the trap axis (Fig.~\ref{Fig:TrapAssemblyAndBeamlines}(a)) providing an unobstructed path for axial lasers exiting the trap. The mesh presents a far-out-of-focus occlusion for topview imaging, and as such does not significantly impact imaging performance. 

\subsubsection{Trap support structure and vacuum assembly}\label{Sec:TrapSystemAssembly}

Functionality of this system, and integration with a horizontal-bore high-homogeneity magnet, motivates design choices enabling delicate structures to be positioned precisely, repeatably, and stably on long cantilever arms that connect the external apparatus with the experimental zone. The design we present here and its mounting to ultra-high-vacuum structures is intended to facilitate trap positioning and suppress potential sources of mechanical instability. This can be especially important as mechanical vibrations are known to deleteriously affect coherence times, imaging quality, and light-matter interaction. 

\begin{figure*}
\centering
\includegraphics[width = 17cm]{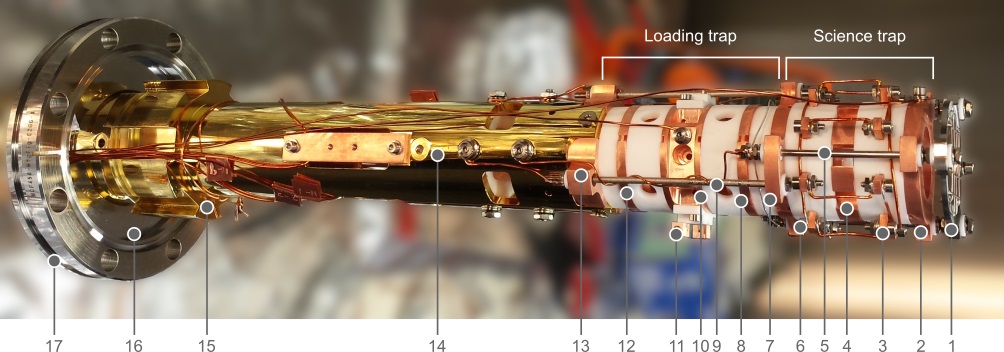}
\caption{Trap assembly mounted on alignment tube (gold), and attached to vacuum flange at bottom.  Assembly is performed in a vertical orientation to ensure proper alignment and to facilitate electrical connections. (1) Wire mesh,  (2, 4, 7) Science trap -- end cap 2, center ring, end cap 1, (3,6) Rotating wall electrodes (vertically-opposite), (5, 9) Titanium rods, (8, 10, 12) Loading trap -- end cap 2, center ring, end cap 1, (11) Beryllium oven, (13) Trap assembly mounting plate, (14) Gold-plated copper alignment tube, (15) Enlarged outer diameter for mating with cuvette neck, (16) Groove grabber, (17) CF63 through flange.}

\label{Fig:trap_tube_assembly}
\end{figure*}

We perform the mechanical assembly of the various trap electrodes and insulators in a clean environment, and integrate additional alignment and mechanical support structures into the assembly.  Fig.~\ref{Fig:trap_tube_assembly} shows the assembled science and loading traps, where the stacked electrodes and Macor rings are clamped together with threaded titanium rods.  The trap assembly is bolted onto a rigid alignment tube, itself clamped at its base in a groove grabber mounted in a CF63 through flange, and further partially inserted into a second support tube inside the main vacuum chamber.   The alignment tube is machined from oxygen-free high-conductivity copper and is gold-plated to avoid cold welding to \eg~trap electrodes or other alignment structures in the presence of snug mechanical fits required to provide stiffness in a horizontal cantilevered orientation. 

We design our vacuum chamber to include a glass cuvette surrounding the trap electrodes, constructed with a glass-metal interface and a short metal nipple for mating to standard ultra-high-vacuum components and provision of stable surfaces for bracing of the trap support structures. The CF63 flange shown on the trap assembly is sandwiched between similar flanges on the main vacuum chamber and the glass vacuum cuvette.  Just above the CF63 flange in the trap assembly, the outer diameter of the alignment tube enlarges to fit snugly with the inner diameter of the stainless steel neck of the cuvette, see (15) and (16) in Fig.~\ref{Fig:trap_tube_assembly}. The outer diameter of this section of the alignment tube is $\emptyset$\SI{54.03}{\milli\meter}, providing a tight H7/h6-class fit with the $\emptyset$\SI{54.00}{\milli\meter} inner diameter of the cuvette's stainless steel neck. Vented channels in the enlarged section allow cabling from the traps to reach the vacuum feedthroughs downstream. 

\begin{figure*}
\centering
\includegraphics[width=17cm]{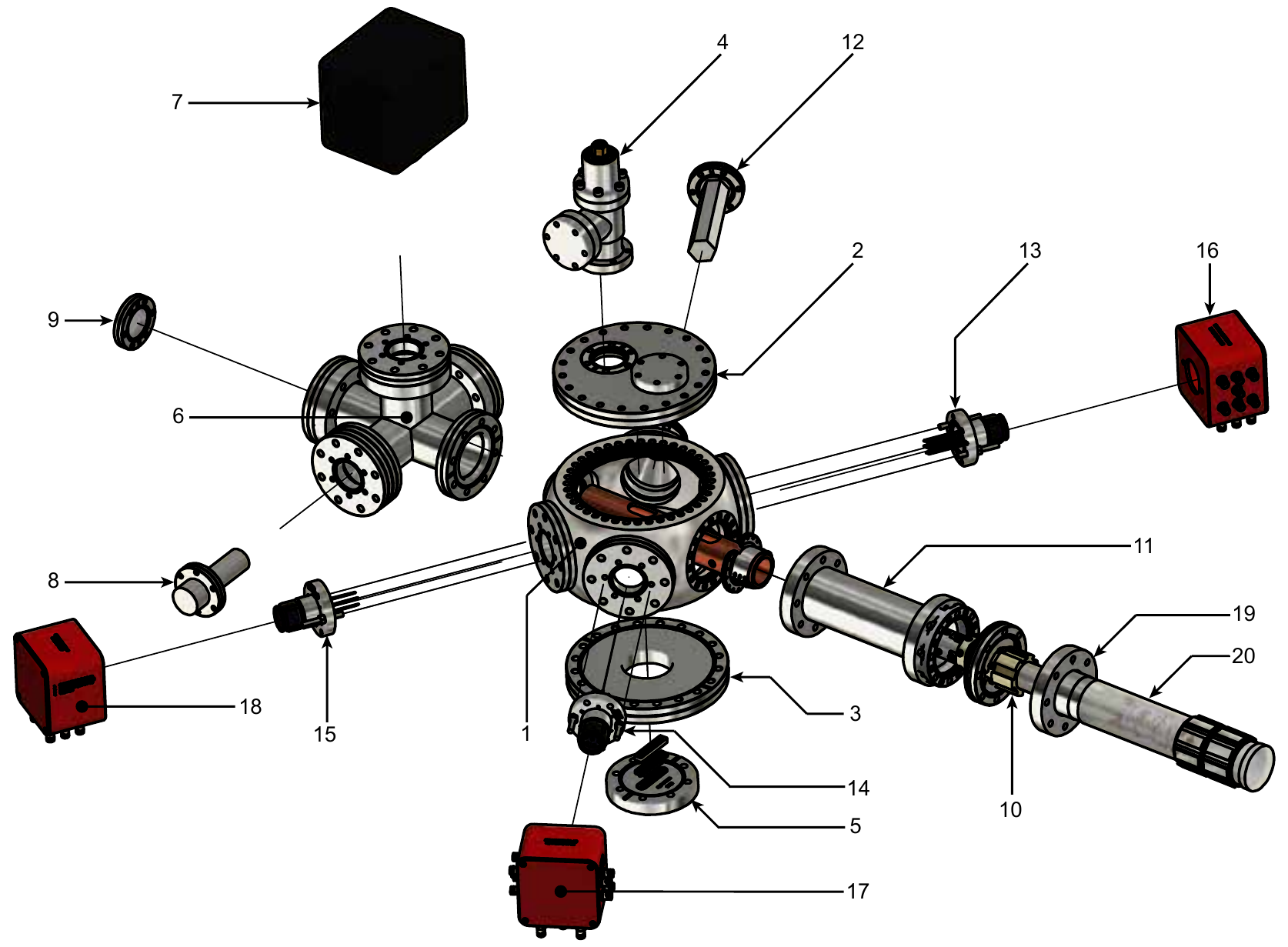}
\caption{Vacuum system assembly. For complete parts list identifying all labels see Table~\ref{Table:vacuum_system_parts_list} in Appendix~\ref{Appendix:UHVPartsList}. Trap tower assembly (10) is encased in a glass cuvette (19, 20) connected to a hexagonal vacuum chamber (1). A pair of CF160 zero-length reducer flanges is mounted on the top and bottom of the hexagon, supporting a CF40 all-metal valve on the top, and a 24-pin CF63 D-Sub ultra-high-vacuum feedthrough on the bottom. A CF63 5-way cross is attached to the hexagon along the axial direction on the back flange, hosting a \SI{20}{\litre\per\second} ion getter pump (7) on top and a Bayard-Alpert ion gauge (8) on the side. The system includes a non-evaporative getter pump (12) mounted on one of the four CF63 side flanges of the hexagonal chamber. The trap and rotating wall voltages are controlled by high-voltage power supplies, routed through high voltage connectors (SHV) on custom breakout boxes (16, 17, 18) associated with respective power feedthroughs (13, 14, 15). Beryllium ovens are driven by power supplies connected to the D-Sub feedthrough (5). After bakeout an ultimate vacuum of order \SI{2e-11}{\milli\bar} was reached, as read off the ion gauge.  In operation, pressure readings increased due to the fringing fields from the superconducting magnet altering the measurement of the pressure gauge and/or pumping efficiency. } 
\label{Fig:exploded_vacuum_assembly}
\end{figure*}

The cuvette itself is designed to provide optically transparent access in a geometry matched to the underlying trap symmetry, see Figs.~\ref{Fig:exploded_vacuum_assembly} and~\ref{Fig:vacuum_assembly_with_cuvette}.  It consists of a \SI{50}{\milli\meter} inner-diameter double octagonal cell, with an eight-fold symmetric array of apertures for radial optical access, each covered with $\SI{15}{\milli\meter}\times\SI{37}{\milli\meter}$ flat rectangular windows externally bonded to the main glass body. The top is sealed with a \SI{60}{\milli\meter} outer diameter window providing axial optical access. All windows are made from UV-grade fused silica, and are anti-reflection coated for \SI{313}{\nano\meter} on both sides. As indicated above, the cuvette neck is mated via a glass-metal interface onto a 316LN stainless steel tube welded to a CF63 flange. 

The trap assembly and cuvette mate to an ultra-high-vacuum system based on a 316L stainless steel hexagon with six CF63 flanges arranged around the sides, and two CF160 flanges on top and bottom.  The trap assembly and cuvette are connected to this hexagon via a CF63 full nipple, and an anti-reflection coated viewport is employed on the back side for axial optical access.

The remainder of the vacuum system consists of a number of standard vacuum components, two permanently attached vacuum pumps, and a number of electrical feedthroughs, see Fig.~\ref{Fig:exploded_vacuum_assembly}. All standard vacuum components are made from 316LNS stainless steel, and custom parts from either oxygen-free high-conductivity copper, grade 2 titanium, or Macor, satisfying the dual requirements of low magnetic permeability and vacuum compatibility. Further details appear in Fig.~\ref{Fig:exploded_vacuum_assembly}.

The side flanges of the vacuum chamber are equipped with 7-pin power feedthroughs for delivering high voltage to the loading trap, science trap, and rotating wall electrodes. Kapton-coated wires are used to connect internal feedthrough pins to all trap surfaces, electron guns (not used or discussed further), and beryllium ovens. A perforated $\emptyset\SI{50}{\milli\meter}$ oxygen-free high-conductivity copper tube is mounted inside the hexagon along the trap axis via internally attached groove grabbers to prevent feedthrough cabling from obstructing optical access along the trap axis and to stabilize the trap alignment tube further.  For full details on component designs see Ref.~\cite{HBall2018PhD}.

%
\subsection{Superconducting magnet and dual-stage reliquefier}
\label{Sec:MagnetReliquefier}

\begin{figure}[b]
		\includegraphics[width=8.5cm]{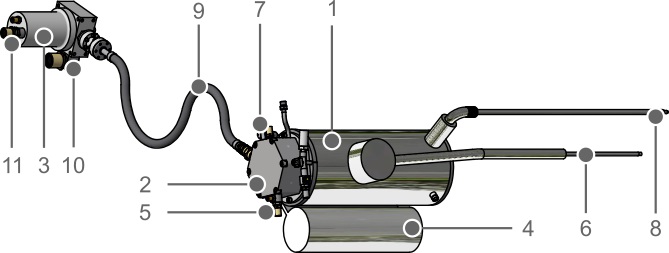}
	\caption{Reliquefier and remote motor assembly (shown \SI{90}{\degree} rotated): (1) Vacuum-sealed condensing chamber, (2) base plate for pulse-tube cold head, (3) remote motor assembly, (4) external helium reservoirs, (5) nitrogen gas bleed inlet, (6) LN$_2$ return line, (7) helium gas recovery inlet, (8) LHe return line, (9) \SI{1.5}{\meter} flex line connecting pulse tube to remote motor assembly, (10) Aeroquip fitting for low-pressure flex line, (11) Aeroquip fitting for high-pressure flex line.}
	\label{Fig:reliquifier_schematic}
\end{figure}

In a Penning-trap system a superconducting solenoid magnet is often used to provide radial confinement of charged particles, and to establish a stable electromagnetic environment for precision experiments. Such experiments include, for example, high-precision mass spectrometry \cite{marshall1998fourier, myers2013most}, determination of fundamental constants~\cite{Hanneke2008,Sturm2014,Heisse2017}, CPT tests \cite{Ulmer2015,Smorra2017}, and quantum simulation with large ion crystals \cite{Britton2012,BohnetScience2016}. These capabilities arise from the high temporal stability and self-shielding effects of superconducting solenoid magnets operated in persistent mode~\cite{Gabrielse1988}.  

In our implementation we employ a homogeneous magnetic field of $B_0 \approx \SI{2}{\tesla}$, in a $\SI{150}{\milli\meter}$-diameter horizontal-bore superconducting magnet.  It is charged below its design field strength of \SI{3}{\tesla} such that the resulting qubit frequency remains in a band where commercial, low-phase-noise, high-power, millimeter-wave sources are available.  By using additional so-called shim coils, spatial field inhomogeneities originating from the geometry and mechanical tolerances of the main solenoid coil can be compensated, resulting in a dedicated region at fields of several Tesla with high field homogeneity and a spatial extent of a few cubic-millimeters. 

One drawback of such systems is the generic need for cryogenic operation of the superconducting solenoid.  Our system includes both a liquid helium (LHe) vessel, and a liquid nitrogen (LN$_2$) shroud to prolong the lifetime of the cryogens.  The magnet cryostat boils off LN$_2$ at a rate of \SI{7.45}{\litre/\day} and LHe at a rate of \SI{0.48}{\litre/\day}. With vessel capacities of about \SI{32}{\litre} for LN$_2$ and about \SI{75}{\litre} for LHe, the \SI{50}{\percent} evaporation times are about \SI{2.2}{\day} and \SI{156}{\day}, respectively.   Despite the relatively long hold time of LHe, frequent refill cycles for LN$_2$ reduce the maximum duty cycle of an experiment.  This is because the magnetic field will change during the filling process due to permittivity variations of the material in thermal contact with the cryostat, as well as pressure changes in the cryostat. The time for the magnetic field to resettle varies for each individual system and on environmental conditions, but can be of order hours. 

To reduce cryogen losses and minimize the impact of cryogen refilling on system up-time, a novel dual-stage reliquefier (DSR) system is used to recondense both helium and nitrogen boil-off, and return the reliquefied gas directly to the cyrostat. This hybrid near-zero-loss system can also operate for several days without power in free-running mode, in contrast to fully cryogen-free systems.  As we demonstrate below, DSR operation has had limited observed impact on the temporal stability of the magnetic field or mechanical vibrations in the trap itself.  Here we provide an overview of system configuration and performance measured over an approximately 12 month period.

The DSR system consists of a two-stage cryocooler pulse-tube cold-head (Cryomech PT407-RM) and a \SI{8}{\kilo\watt} helium compressor (Cryomech CP2870). The pulse-tube cryocooler is mounted inside the vacuum-sealed condensing chamber of the reliquefier. The first stage of the cold-head provides cooling capacities at the temperature range of \SI{30}{\kelvin} to \SI{70}{\kelvin} for recondensing LN$_2$ boil-off with specified cooling power of \SI{37}{\watt} at \SI{65}{\kelvin}. The second stage provides cooling capacities of \SI{0.5}{\watt} at \SI{4.2}{\kelvin} for recondensing LHe boil-off.  The main reliquefier assembly is shown in Fig.~\ref{Fig:reliquifier_schematic} and has been customized for our application; for detailed schematics and performance specifications see Ref.~\cite{WangOIP2015}. The associated compressor package uses \SI{99.999}{\percent} pure helium as the refrigerant gas and is installed in the service room adjacent to the lab, and is decoupled from the floor using vibration-dampening feet to suppress low-frequency vibrations.

The DSR is mounted above the magnet on a gondola structure decoupled from the walls and floor of the laboratory to mitigate the coupling of vibrations into the magnet. One key feature of the cold-head design is the separation of the remote motor assembly from the pulse-tube. The remote motor controls a valve which sets the pressure changes inside the pulse-tube, creating the necessary cooling power to achieve temperatures below \SI{4}{\kelvin} at the heat exchanger.  The physical separation of this unit from the cold-head via a \SI{1.5}{\meter} flex line permits a helium expansion cycle in the cold-head without directly using a displacer or piston, thereby reducing vibrations in the magnet which are known to limit the trapped-ion spin coherence times \cite{BrittonMagnet2016}.  The remote motor is also mechanically fixed to the overhead gondola system and sits on a sliding shelf which permits strain management in the flex line during reliquefier insertion or removal.  

Two drain legs of the reliquefier are inserted into fill ports in the magnet and make primary physical contact to the magnet cryostat through elastomere O-ring connections.  Helium boil-off from the magnet cryostat enters the condensing chamber via a flexible vapor line connecting the LHe vessel to a gas inlet on the DSR. The vapor line is equipped with a needle valve setting the helium flow rate and hence permitting control over the pressure in the magnet cryostat. Helium gas entering the chamber contacts the \SI{3.9}{\kelvin} cold head on the second stage of the cryocooler where it condenses. The liquefied helium is then funneled into a \SI{12.7}{\milli\meter}-diameter vacuum-insulated return line inserted into the helium fill port of the magnet cryostat, forming a closed helium loop. For the nitrogen cycle, the return line consists of a coaxial tube; Nitrogen boil-off flows up the outer jacket towards the condenser surface, and the recondensed liquid flows back through the center tube.  This geometry permits both capture of LN$_{2}$ boil-off, and its return after recondensing through the same port.

The differential pressure between cryostat and laboratory, as well as the temperature at the two cooling stages, are monitored by sensors. A stable cryo-siphon loop is achieved when the liquefication rate is slightly lower than the boil-off rate. The helium cycle is regulated at a temperature of \SI{3.9}{\kelvin}, measured at the cold head. By setting the needle valve mentioned above, a stable overpressure of about $1\,\mathrm{psi}$ in the helium cryostat can be maintained. The nitrogen cycle is regulated at an overpressure of $0.4$ to $0.5\,\mathrm{psi}$, resulting in a temperature of about \SI{64}{\kelvin} at the first stage of the cold head. Stable operation is achieved using a feedback controller (Stanford Research Systems CTC100) to regulate heaters at the nitrogen and helium condensing stages. Maintaining a positive pressure in the cryostat vessels is imperative since atmospheric gases can freeze inside, and subsequently clog the two cryogenic vessels. 

Manually refilling LHe or LN$_2$ is not possible with the reliquefier installed as it occupies the relevant liquid-cryogen fill ports. A top-up of cryogens to compensate residual system losses is achieved by slowly bleeding gaseous helium and nitrogen into the recondenser. Helium gas is injected via a flange on the magnet cryostat's helium manifold connected in tandem to the helium vapor line of the reliquefier. Nitrogen gas is bled in directly through an inlet on the condenser chamber.  As such, the systems's liquid cryogen levels may be maintained continuously without the need to interrupt system performance.  

We measure the effective loss rate of cryogens from our system by calculating the total input gas volume that becomes liquefied using a digital flow meter (Bronkhorst MV-392-He) installed on the helium inlet line.  Once a sufficient amount of helium gas is liquefied, the rising LHe contacts warmer system components at the top of the vessel and results in an increase in boil-off rate. The helium pressure therefore increases in the magnet cryostat, reducing the flow rate of injected gas to zero. Integrating the flow rate up to this point, and converting from gas to liquid volume yields the estimated volume of LHe added to the system. Using this method we calculate the total volume of added LHe after 77 days of continuous operation to be \SI{0.6}{\litre}, or a loss of \SI{7.8}{\milli\litre/\day}. Over a year this implies a loss of less than \SI{5}{\percent} of the helium vessel capacity. The loss rate for liquid nitrogen determined from the level meter is \SI{30}{\milli\litre/\day} (this value is strongly dependent on the pressure setpoint in the feedback system, likely through leakage in the non-return valves). Thus, the loss rates for LHe and LN$_2$ are decreased by more than a factor 60 and 240, respectively. This results in a time-to-empty of about 9600 days for LHe and 1060 days for LN$_2$.  We have periodically repeated the top-up procedure over more than a year of continuous operation and found the loss rate to be consistent.  All sealings between the DSR and the cryostat are compression fittings using FKM-rubber (from German Fluorokarbonmaterial) O-rings. The cryostat features non-return (pop-off) valves and a burst disc to permit emergency decompression of the vessels in the event of a superconducting quench, which likely constitute the largest component of the overall system leak rate.

The benefits of pulse-tube cryocoolers are balanced against potentially deleterious performance limitations arising from vibration-induced magnetic field fluctuations~\cite{Brown_RSI, BrittonMagnet2016}.  In order to characterize the impact of the DSR on magnet performance, the homogeneity of the magnetic field in the center of the bore, and its temporal stability with the DSR running were mapped using a commercial NMR probe. Such systems are readily available, and are regularly used to characterize magnet performance. The NMR system used in this work consisted of a probe (300/89, General Electrics) with modified circuitry to tune to the required frequency, a console (Redstone, tecmag), and a radio-frequency power amplifier (Tomco Technologies). Free induction decay measurements were performed on a sample consisting of an aqueous \SI{0.15}{\mol/\litre} copper sulphate solution. 

\begin{figure}[t]
	\centering{
		\includegraphics[width=\columnwidth]{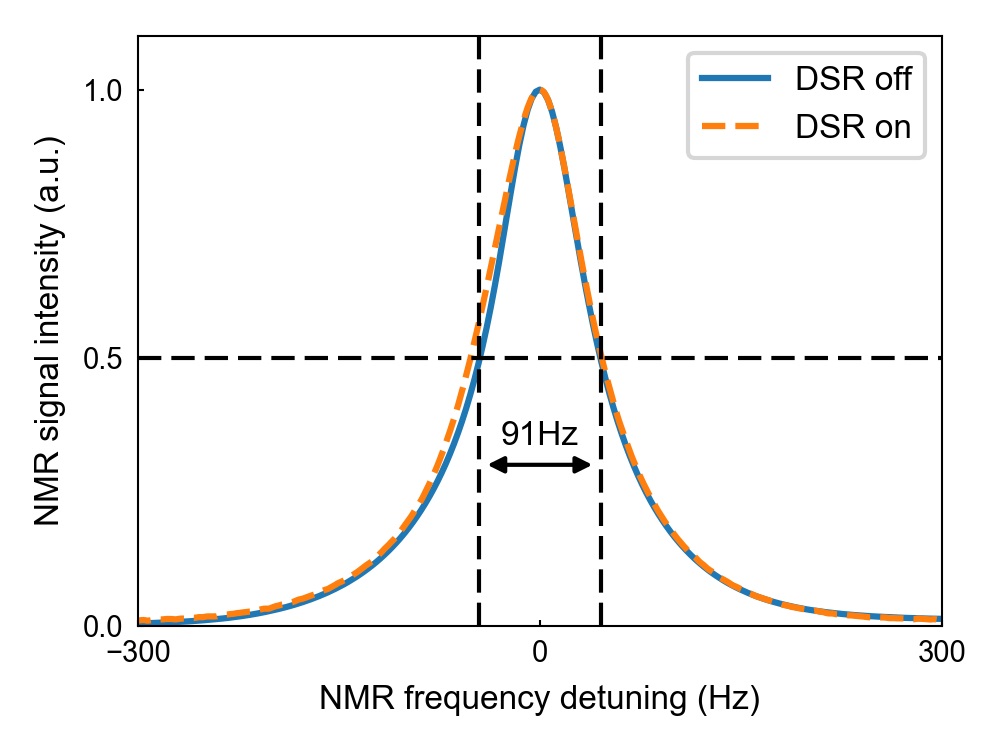}}
	\caption{Comparison of the nuclear magnetic resonance lineshape as a function of detuning from $\nu^{\text{NMR}}=\SI{85.064094}{\mega\hertz}$ at the nominal ion position with the dual-stage reliquefier turned on and off. No significant change could be observed at this level of sensitivity.  The $\emptyset 7.0 \times$\SI{8.0}{\milli\meter^3} cylindrical sample was contained in a nylon capsule mounted inside the probe, which was mounted inside the bore in a translating and rotating structure, providing fine control over the sample position. }
	\label{Fig:ShimCoilCharacteristics}
\end{figure}

The NMR signal's optimal full-width-half-maximum (FWHM), with the DSR off, and following standard shimming procedures, was $\Delta\nu_{\mathrm{FWHM}}^{\mathrm{NMR}}=\SI{91}{\hertz}$ at a Larmor frequency of $\nu^{\mathrm{NMR}}=\SI{85.064094}{\mega\hertz}$.  This measurement was repeated with the DSR turned on. The resultant resonance lineshapes are compared in Fig.~\ref{Fig:ShimCoilCharacteristics}.  We observe minimal deviation between the two cases, thus indicating no significant increase in magnetic field inhomogeneity due to the operation of the DSR at the resolvable scale.  
The measured linewidth indicates a fractional inhomogeneity of approximately 1\,ppm, which corresponds to about \SI{7.6}{\milli\meter} diameter spherical volume.  The measured NMR linewidth is broadened due to the concentration of the copper sulfate solution, and this figure represents a likely upper bound on field inhomogeneities. 

The measured NMR center frequency also helps identify the operating magnetic field  of $B_0\approx\SI{1.99793}{\tesla}$.  Due to losses in the superconductor, the field intensity in the bore slowly decreases over time. To quantify this the center frequency of the resonance signal was measured over a 24-hour period, resulting in a measured drift rate of $\mathrm{d}B_0/\mathrm{d}t = \SI{-3e-8}{\tesla\per\hour}$. This ultimately yields \SI{}{\kilo\hertz} level drifts in the qubit frequency over the course of an hour, orders of magnitude smaller than the achievable Fourier-linewidth of a driven spin transition as previously reported in e.g.~\cite{Britton2012}.

It is important to note that these measurements were performed in a magnetic environment that is different to the one the ions experience; notably lacking all optomechanics, the vacuum chamber, and the trap itself. The presence of these largely metallic structures can influence the magnitude and spatial dependence of present magnetic field distortions. A more direct and precise in situ measurement of magnetic field inhomogeneities will be discussed in section~\ref{Sec:Conclusion}.

\subsection{Inbore optomechanics and imaging system}
\label{Sec:Optomechanics&ImagingSystem}

In this section we describe the custom inbore optomechanics used to route laser light to and from the trapping region, deliver millimeter waves to the ions, and capture the ionic fluorescence used for diagnostics and qubit-state readout (Fig.~\ref{Fig:SetupOverview}).  This subsystem joins in a non-contact fashion with the trap at the center of the magnet bore and also connects to external optical systems. Below we introduce key design elements of the optomechanical assembly, and provide detailed characteristics of the custom imaging system designed for site-resolved topview ion imaging.

\subsubsection{Optomechanical system design}
The design of this subsystem is constrained by the geometry of the trap and the magnet-bore diameter. All components must fit within the $\SI{150}{\milli\meter}$ diameter horizontal bore, and components which are radially aligned to the trap must fit within a radial shell constrained centrally by the \SI{68}{\milli\meter} outer diameter of the trap cuvette.  The construction must also permit independent adjustment of all pertinent degrees of freedom in order to ensure good optical alignment with trap apertures, as well as positional repeatability. The optomechanics further face the stringent requirement of high mechanical stability to ensure lasers remain aligned to the ion cloud of order microns over long experimental runs, and must be sufficiently rigid to maintain good alignment while the assembly is inserted into the magnet bore. More stringent requirements still hold for alignment of the Raman beams to generate an optical dipole force for quantum simulation.

These requirements are further complicated by the limited mounting options arising from the system geometry.  The assembly may only be mechanically anchored outside of the magnet bore, meaning it must either be cantilevered inside the bore or stabilized against the inner diameter of the bore itself. Moreover components must exhibit low magnetic susceptibility to prevent local field distortions near the sensitive experimental zone, this also excludes nominally non-magnetic austenitic steel variants such as 316 for inbore construction. 

\begin{figure*}%
\begin{center}
\includegraphics[width=17cm]{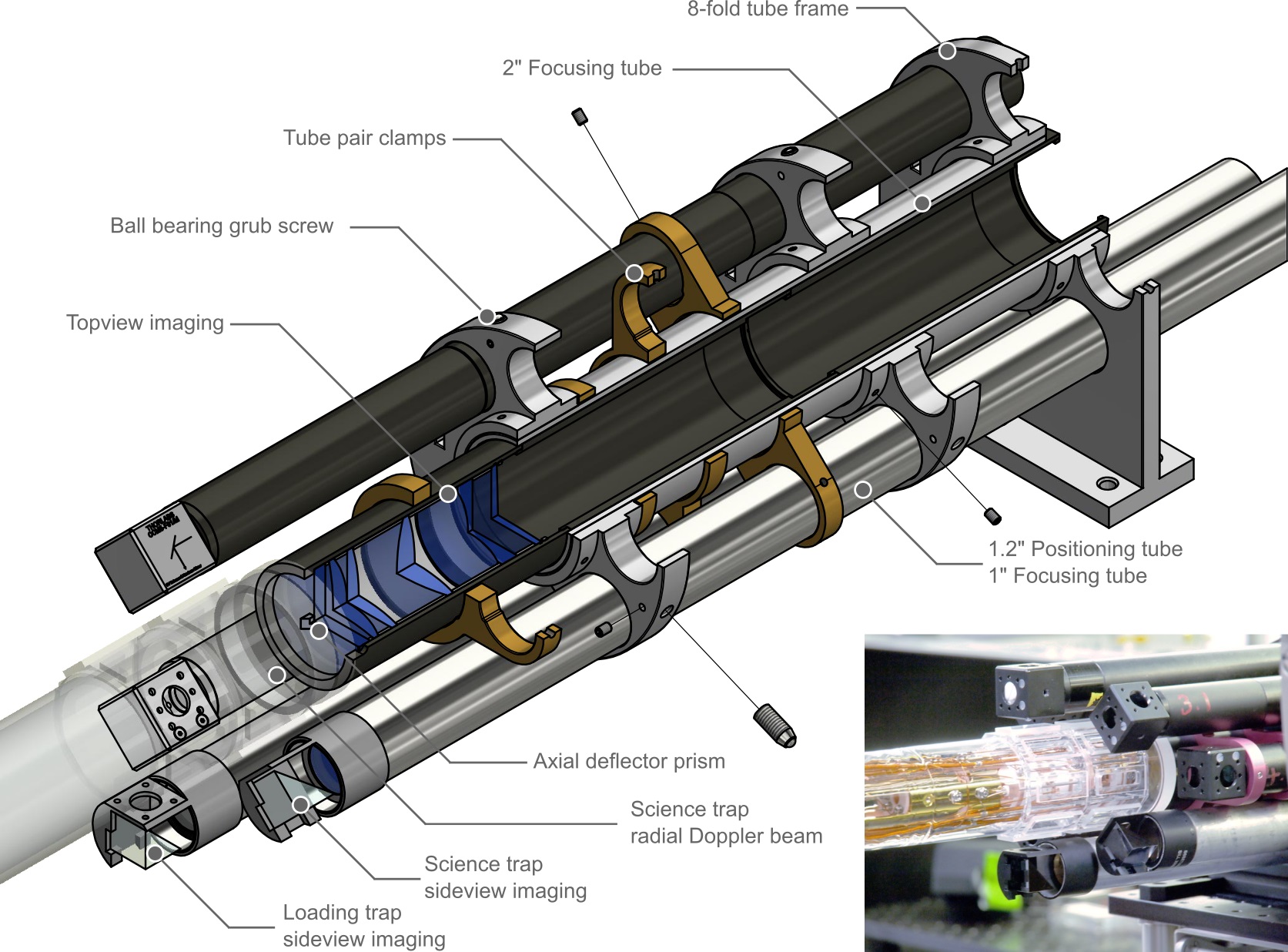}
\caption{Section view of inbore optomechanics.  Assembly of eight-fold tube frame with captured ball bearing screws and fastening grub screws indicated. Note that three of the seven tubes are not shown and the millimeter wave parts are omitted. Inset: Photograph of trapping and optomechanics sections joined extrabore during pre-alignment.}
\label{Fig:Inbore_Optomechanics}
\end{center}
\end{figure*}

\begin{figure*}%
\begin{center}
\includegraphics[width=17cm]{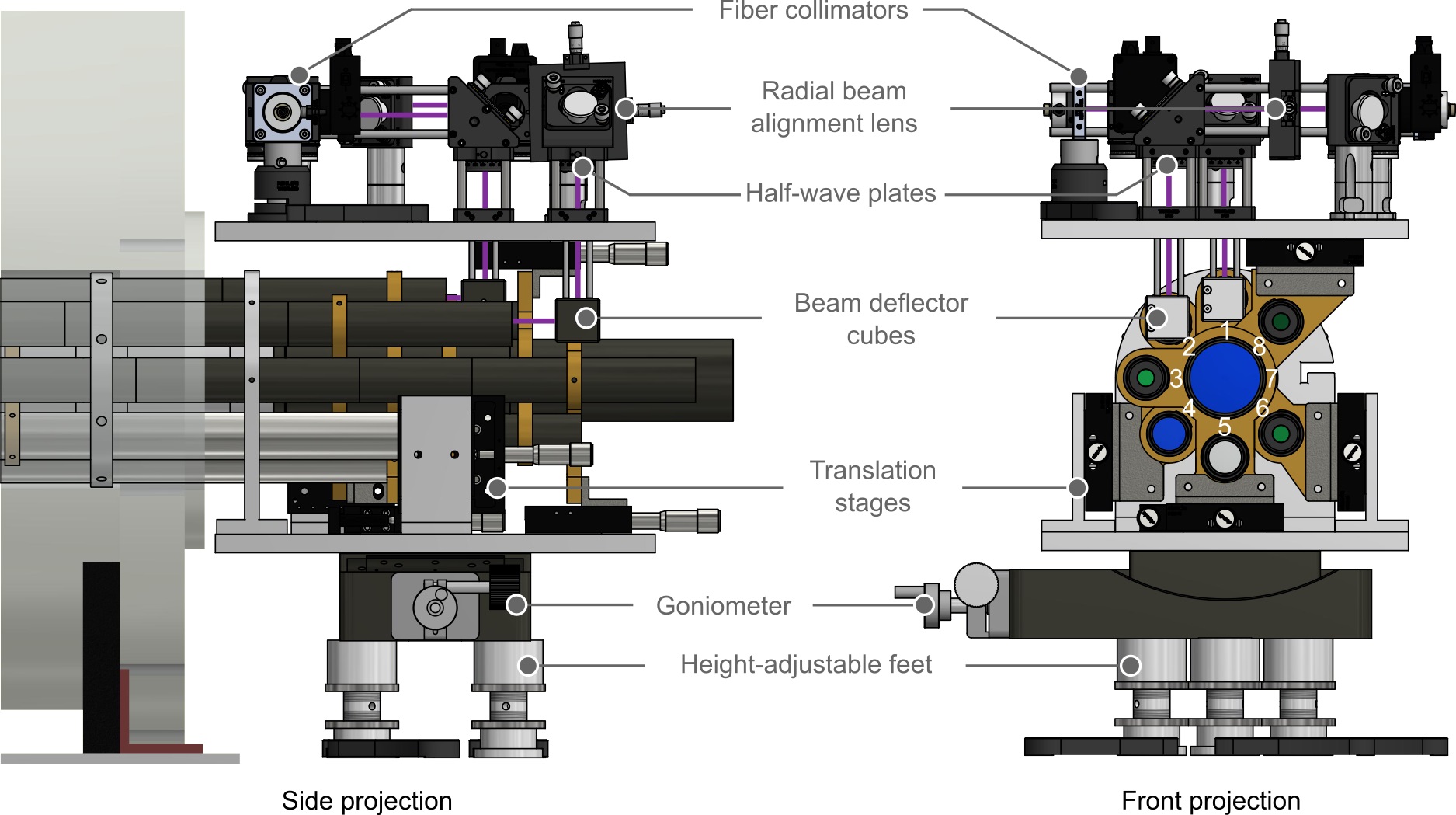}
\caption{Side and front projection view of extrabore optomechanics. Purple indicates propagation of Doppler cooling beams. Light leaving the optomechanics, supporting posts for top platform, and magnet in front projection omitted for clarity. Eight-fold positions are: (1) Loading trap radial Doppler beam entry, (2) science trap radial Doppler beam entry, (3) axial beam exit, (4) science trap sideview imaging, (5) loading trap sideview imaging or radial Doppler beam exit, (6) science trap radial Doppler beam exit, (7) millimeter waveguide port, and (8) science trap sideview alignment port or additional sideview port.}
\label{Fig:Extrabore_Optomechanics}
\end{center}
\end{figure*}

Our system (Fig.~\ref{Fig:Inbore_Optomechanics}) is based on a central, eight-fold symmetric frame that matches the symmetry of the underlying trap electrodes.  The core mechanical components are a central frame tube and two inbore discs affixed using radially oriented grub screws.  This tube also attaches to an external plate that fixes the frame to the extrabore optomechanics (Fig.~\ref{Fig:Extrabore_Optomechanics}), and possesses an inner diameter machined to allow insertion of a standard 2" optics beam tube.  The three discs feature eight radially arranged cutouts that allow insertion of five standard 1" beam tubes, two 1.2" beam tubes, and one WR-19 waveguide.

The assembly is approximately self-centering in the magnet's bore by means of captured, spring-loaded ball-bearing grub screws inserted radially into the two discs. These are adjusted in height prior to assembly of the full frame from the inside of the discs, and fixed in position by additional fastening grub screws as indicated. All grub screws used in the assembly are grade 2 titanium.

The seven tubes affixed to the inbore discs are terminated at the trap with deflector prism mounts.  The use of each tube is indicated in Fig.~\ref{Fig:Extrabore_Optomechanics} and ranges from beam delivery to the ions, to positioning of imaging elements.  Tubes are arranged in pairs and fixed in orientation by means of tube-pair clamps.  The clamps maintain pairwise alignment while the tubes are adjusted with extrabore translation stages. The tubes and clamps are fixed by nylon-tipped grub screws made from grade 2 titanium, opposite extrusions that guarantee good line contact between clamp and tube. All positioning features were machined to H7/h6 locating transition fits with medium interference that allow repeatable positioning without undue pressure on extrabore actuator stages. 

Deflectors employed for radial Doppler beams in either trap are Thorlabs CCM5 cubes with custom, high-reflectivity right angle prisms and nylon screws. The two imaging tubes feature a custom prism mount that does not limit achievable numerical aperture, and positions the prisms more accurately. Imaging tubes consist of an outer 1.2" Thorlabs SC1800RL dust cover onto which the prism holder is press-fitted and glued, and an inner 1" beam tube. The outer tube allows optimization of the deflector's axial alignment with the trap, while the inner tube permits independent adjustment of the image focus.  

The millimeter-wave routing takes the remaining slot in the eight-fold-symmetric frame and features a loose transition fit cutout for WR-19 waveguides, see position 7 in Fig.~\ref{Fig:Extrabore_Optomechanics}. The waveguide and custom delivery system are restrained by another ball-bearing grub screw on the last tube clamp, joining it to the 2" tube perpendicular to the trap axis, but leaving axial translation unrestricted. 

The central 2" beam tube that houses the topview imaging system is joined with one of the five 1" beam tubes by means of a clamp. An axial deflector \SI{5}{\milli\meter} right-angle prism is glued onto the first element of the imaging system to deflect axially propagating laser beams away from light-sensitive cameras and photomultiplier tubes. The prism is located behind the retaining lip of the imaging system housing which prevents contact of the sharp edge of the prism with the cuvette window. The deflected light travels down the associated 1" tube and is used for alignment diagnostics of the axial beams outside of the magnet bore. 

The inbore assembly is connected to an aluminium breadboard mounted on a custom goniometer as shown in Fig.~\ref{Fig:Extrabore_Optomechanics}. The goniometer allows global rotation of the optomechanical assembly coaxial with the magnet's bore in order to compensate for rotational misalignment to the trap. It is mounted on height-adjustable feet with fine-pitch threads to set the vertical position of the assembly. This also allows adjustment of the optomechanical system's pitch relative to the trap axis.  Non-magnetic linear translation stages are mounted on the breadboard and connected to the tube-pair clamps allowing individual axial translation of these tubes.  The tubes employed for sideview imaging in the science trap have two translation stages to allow individual fine positioning and focusing. A second horizontal platform on top of the goniometer (mounting posts not shown) sits above the bore which hosts fiber collimation and beam steering optics.

All laser light is delivered to the trap via solarization-resistant, UV-cured, polarization-maintaining fiber patch cords\cite{MarciniakOSA2017} followed by beam-shaping optics,  mounted externally to the magnet bore (Fig.~\ref{Fig:Extrabore_Optomechanics}).  The use of fibers for this purpose provides robust beam pointing stability as well as spatial and polarization filtering, while also decoupling the alignment of the optomechanical system from external laser systems. The fibers are connected to collimators which are purged with dry N$_2$ gas to prevent UV-induced accretion of particles onto the fiber facet. 

The primary laser beam delivered by the inbore optomechanical system is the radial Doppler cooling beam.  Light from the collimator passes beam shaping optics whose final element is a long-focal-length lens in a $x$-$y$-translation mount. The lens both focuses the radial beam to a waist of $w_0 \approx \SI{45}{\micro\meter}$, and allows pure translation of the waist in the direction parallel and perpendicular to the trap axis. As a consequence of the arrangement of deflectors the beam translation axes are rotated by \ang{45} with respect to the translation mount. An adapter plate (not shown) can be used to rotate the translation mount to account for this. After the translation mount the beam passes a half-wave plate and is then deflected down the Doppler beam tubes using the same CCM5 deflector cubes referenced above.

Axial beams (delivery not shown) for repump, cooling and photoionization lasers are delivered in a similar fashion via the vacuum chamber viewport. Fiber-delivered \SI{313}{\nano\meter} beams are superimposed on a polarizing beam splitter cube and passed through a spatial filtering setup before they are overlapped with the free-space delivered photoionization beam. We employ 250 TPI pitch micrometer screws and a long-focal-length steering lens to align the axial beams. The diameter of the axial beams inside the trap can be controlled between $w=\SI{0.2}{\milli\meter}$ and $w=\SI{1}{\milli\meter}$ at the position of the ions or the deflector prism by moving the spatial-filter output lens. Similarly, the size of the photoionization laser can be varied around its nominal $w=\SI{1}{\milli\meter}$. 

\subsubsection{Diffraction-limited imaging system} 
\label{Sec:Imaging}

\begin{figure*}%
\begin{center}
\includegraphics[width=17cm]{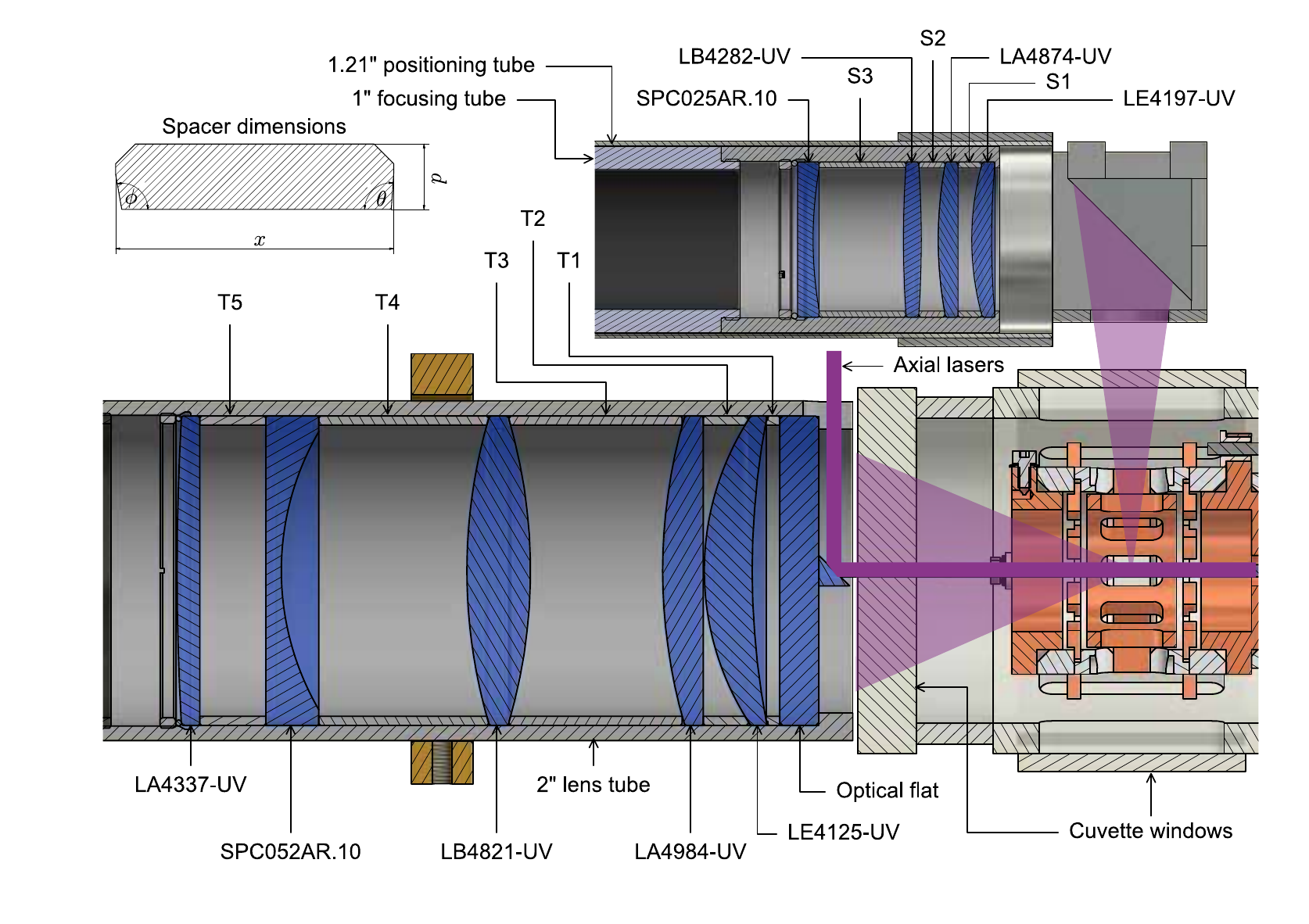}
\caption{Section through topview and sideview imaging systems around the ion trap. The relative orientation of the section planes in both views has been altered for clarity. Topview lens spacers are labeled T1 - T5, sideview lens spacers are labeled S1 - S3. The extrabore imaging sections with spacer T6 and the imaging plano-convex lens are not shown (see text). Lenses starting with L are Thorlabs stock items, lenses starting with SPC are Newport stock items. Spacer parameters are ring thickness $d$, distance $x$ between lens contact points, and contact angles $\theta$ and $\phi$. Both imaging systems are secured within lens tubes by a retaining ring pushing onto an O-ring to avoid scoring, or breaking the lenses upon fastening. The deflector prism minimizes light leakage by including an optically flat Si wafer on the front side which has high reflectivity in the UV and is sufficiently thick to prevent transmission.}
\label{Fig:Imaging}
\end{center}
\end{figure*}

In this subsection we describe the optical imaging system designed and constructed to enable the detection of ions in the Penning trap. Diffraction-limited, site-resolved imaging of large crystals is a well demonstrated capability in Paul traps\cite{kaufmann2012,bermudez2012,thompson2015ion}, but becomes challenging in Penning traps due to the ions' magnetron motion. Groups at NIST and Imperial College London have demonstrated the capability to perform diffraction-limited imaging in the radial line of sight, see e.g. Ref.~\cite{mavadia2013}, and diffraction-limited, site-resolved imaging of large ion crystals in the axial line of sight has been demonstrated at NIST, see e.g. Ref.~\cite{Mitchell1998,BiercukQIC2009}. However, to the best of our knowledge a detailed presentation of the design and performance characteristics of these systems is lacking in the literature.

The system we present to achieve similar simultaneous imaging is constructed in two primary parts; topview imaging along the trap axis allows site-resolved ion imaging, and sideview imaging along a radial direction permits simplified diagnostics, determination of ion crystal dimensionality, and global fluorescence measurements.  Overall we target high-numerical-aperture systems in order to allow spatial resolution with ion-ion distances of approximately \SI{10}{\micro\meter} and state discrimination on tens to hundreds of microseconds timescales.  An additional simplified imaging system was installed on the sideview port of the loading trap to monitor \SI{235}{\nano\meter} atomic fluorescence along with ionic fluorescence on a photomultiplier tube, but does not provide diffraction-limited imaging.

Our approach to the design of this system incorporates generic considerations such as tolerance to misalignment of optical elements.  However, we also encounter a number of challenges posed by the structure and geometric constraints of our system:

\begin{enumerate}[i]
    \item The vacuum cuvette windows form the first optical element in either path where they set a minimum working distance and add aberration.
    \item A UV-opaque deflector prism is mounted on an optical flat between the cuvette and topview imaging optics to prevent axial lasers from striking the topview camera.  However, this produces a rectangular obstruction in the center of the image and sets a minimum offset between the cuvette flat and imaging optics.
    \item Co-location of topview and sideview imaging systems in the magnet bore constrains the size of available lenses.
    \item The numerical aperture for sideview imaging is limited by the cuvette window geometry.
    \item Active (electronic) imaging elements must be positioned far from the fringing field of the magnet requiring an imaging path of at least one meter.
   \item We target designs using only relatively small numbers of exclusively stock lenses, and easy adaptability to small changes in operating wavelength for flexibility and cost efficiency.
\end{enumerate}

Our system design  is shown in Fig.~\ref{Fig:Imaging} with respective orientations and distances to scale.  For our design magnifications of $\times$30 and $\times$13 (topview and sideview), ion crystals containing hundreds of ions (depending on crystal conformation and ion spacing) can be imaged onto available camera chips within the respective fields of view of \SI{275}{\micro\meter} and \SI{550}{\micro\meter}, within which both are above the diffraction limit. We achieved numerical apertures of 0.32, and 0.12  (f/\# 1.56, and 4.17) at working distances of \SI{50.8}{\milli\meter} and \SI{73.9}{\milli\meter}, for topview and sideview systems respectively. These compare favourably with the geometric limitations on the achievable numerical apertures of 0.34 and 0.13.  Characteristics of the lens positions constituting the imaging objectives, and defined by fixed machined spacers, are summarized in Table~\ref{Table:Spacers} in Appendix~\ref{Sec:AppendixImaging}. Production of the spacers and housing, as well as assembly and interferometric testing were performed by Sill Optics GmbH \& Co KG. 

Optical designs were realized using numeric optimization with the initial aim of maximizing the Strehl ratio of the collimating section of the imaging objectives (up until and including SPC025AR.10 and SPC052AR.10, respectively, see  Fig.~\ref{Fig:Imaging}).  This is a convenient measure of the quality of the optical system which compares the peak intensity of the imperfect, i.e. aberrated, optical system to that of a perfect system limited only by diffraction.  The starting point for our optimization was derived from published approaches to designing imaging systems for vacuum chambers~\cite{Alt2012, Pritchard2016}, and lens separations and focal lengths were iteratively adjusted, constrained by available stock selections.  Next, we proceeded to optimize the image-forming elements, again with the aim of maximizing Strehl ratio for a fixed distance to the magnet and fixed magnification.  

The collimated output of the topview imaging is delivered through a beam-shrinking telescope consisting of a long-focal-length plano-convex singlet (Thorlabs LA4337-UV) and a plano-concave singlet (Newport SPC019AR.10, not shown) with lens spacing of \SI{1197}{\milli\meter}. This enables a target image magnification of $\times$30 when used in conjunction with a plano-convex singlet (Thorlabs LA4579-UV, not shown). The target magnification may be adjusted between about $\times 25$ and $\times 40$ by varying the distance between the last two elements while maintaining diffraction-limitation.  Inserting additional flat, normal-incidence optics in the beam-shrinking telescope does not alter the performance characteristics of the imaging system due to the very low convergence angles in that section of the beam. This is a feature of great practical utility since it offers flexibility e.g. with respect to the number and thickness of optical filters or polarizers installed to reduce unwanted scatter from background light. Omission of all extrabore imaging optics additionally produces a diffraction-limited image at approximately the same position at a fixed magnification.

We find that for the sideview system any single plano-convex spherical singlet with sufficiently long focal length can be used for diffraction-limited imaging at any point in the outgoing beam. The residual divergence from imaging an extended source, however, causes the beam to clip progressively on the light-tight enclosures the farther out the imaging lens is placed. The final imaging element for the sideview system was a Thorlabs LA4663-UV, which sets the magnification to about $\times$13. The sideview lenses are housed in a positioning tube that translates axially independent of the prism used to deflect ionic fluorescence, permitting independent alignment and focusing, see Fig.~\ref{Fig:Inbore_Optomechanics}. 

The calculated imaging performance of both systems is described in detail in Appendix~\ref{Sec:AppendixImaging}. We find that the contrast in both systems closely follows the diffraction-limited performance well beyond the target resolution of \SI{10}{\micro\meter} or 50~cycles per mm, limited primarily by the achievable numerical aperture of geometric constraints. For the topview imaging system we find that contrast falls to 0.5 only at 450~cycles per mm or $\approx\SI{1}{\micro\meter}$ resolution at the edge of \SI{550}{\micro\meter} diameter crystals.

Finally, we extrapolate from these optical measures to performance measures of closer physical significance to imaging of trapped ions.  First, we find the photon collection efficiencies of the two imaging systems from their solid angle in the circularly polarized dipole emission pattern are \SI{3.7}{\percent} and \SI{0.27}{\percent} for the topview and sideview, respectively. The transmission through the imaging systems themselves is calculated to be $\approx\SI{95}{\percent}$ for circularly polarized light, accounting for bulk absorption \cite{edwards1966optical} and reflection loss at all surfaces. With typical detector efficiencies of $\approx\SI{20}{\percent}$ in the spectral region of interest we calculate a total collection of \SI{0.54}{\percent} and \SI{0.04}{\percent} post-detector, respectively. Including conservative estimates of losses in fluorescence filters and mirrors, we estimate a photon detection rate of $\approx\SI{100}{\kilo\hertz}$ per cold beryllium ion on a typical topview detector.  



\section{Trap operation and characterization}
\label{Sec:Results} 
In this section we describe the operating conditions employed in our efforts to observe first light in the trap, as well as site-resolved ion crystals. This section begins with a discussion of the trapping potential employed in our experiments, description of ion loading procedures, and presentation of site-resolved images of ion crystals.  We conclude this section with the characterization of mechanical stability of the trap in situ, looking forward to future experiments.

\subsection{Trap potential tuning and configuration}

The electrostatic potential created by the Penning-trap electrode structure deviates from the ideal quadrupolar potential given by Eq.~\ref{Eq:IdealHarmonicTrappingPotential} in Appendix~\ref{Sec:IdealPenningTrap} due to anharmonicities originating from the finite size of the trap, application-specific modifications, and mechanical tolerances. For instance, in the loading trap the potential in the trapping region is affected by the finite size of the hyperbolic electrodes as well as the apertures present for laser beam access and ion shuttling. 

Deviations from the ideal Penning trap are more significant in the science trap, due to the optical apertures and cylindrical electrode geometry. Accordingly we apply a constant electric potential $V_\mathrm{STRW}$ to the science trap rotating wall electrodes (ST RW), in addition to the rotating wall radio-frequency drive $V_\mathrm{W}$, in order to suppress anharmonicities in the trapping region. 

We perform numeric simulation using the package SIMION~\cite{Dahl2000} in order to find the optimum voltage tuning ratio $T=V_\mathrm{STRW}/V_\mathrm{STCR}$, between the science trap rotating wall DC and science trap center ring electrode (ST CR) voltage. First, the Laplace equation for the complete electrode structure is solved on a 3D anisotropic grid with \SI{100}{\micro\meter} spacing in both radial directions, and \SI{5}{\micro\meter} spacing in the axial direction. Second, the equation of motion is solved for ions starting with different axial and zero radial kinetic energy in the electrostatic center of the trap for a variety of tuning ratios $T$. In an ideal quadratic potential the axial oscillation frequency $\nu_z$, Eq.~\ref{eq:axialfreq}, is independent of the ion's kinetic energy and therefore its motional amplitude. To obtain the axial frequency (and hence calculate anharmonicity), the time-of-flight $t_n$ for an ion to perform $n$ oscillations is computed, giving $n/t_n=\nu_z(T,E_\mathrm{kin})$. This analysis incorporates the effect of all orders of anharmonicities (see Appendix~\ref{Sec:IdealPenningTrap}), and is not limited to the evaluation of the lowest order contributions to anharmonicity as is typically used. 

\begin{figure}[t]
		 \includegraphics[width=8.5cm]{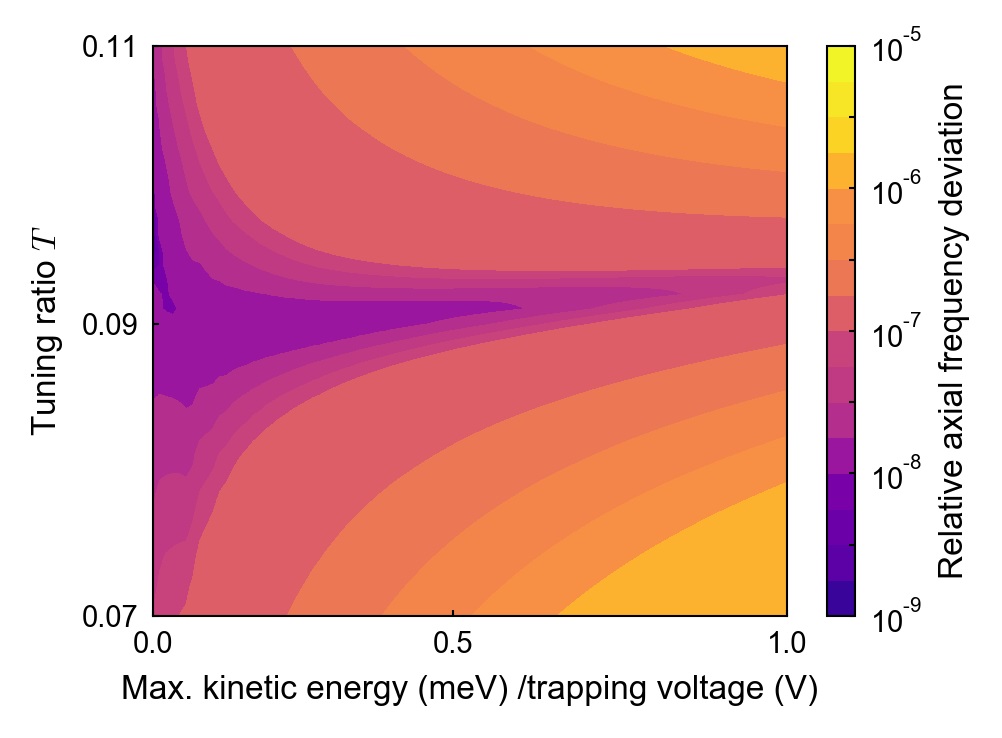}
    \caption{Simulated relative axial frequency deviation as a function of the maximum kinetic energy of an ion ensemble, scaled to the trapping voltage $\lvert V_T \rvert $, calculated for different tuning ratios $T$. At a tuning ratio of about $T_\mathrm{opt}=0.09$, the deviation is smallest over a wide energy range. The irregular structure at small energies is due to numerical inaccuracies in the simulation. For further details see text.}
    
    \label{Fig:SIMION_ST_ax_freq}
\end{figure}

The relative axial frequency deviation has been calculated as the standard deviation of the axial frequencies of an ion ensemble with kinetic energies, scaled to the trapping voltage $V_T= V_\mathrm{STCR} - V_\mathrm{STEC1,2}$, between $E_\mathrm{kin,min}/\lvert V_T \rvert=\SI{0.03}{\milli\electronvolt/\volt}$ and $E_\mathrm{kin,max} /\lvert V_T \rvert=$\SIrange{0.05}{1}{\milli\electronvolt/\volt}, normalized to the axial frequency at minimal energy, $\nu_z(T,E_\mathrm{kin,min})$, shown in Fig.~\ref{Fig:SIMION_ST_ax_freq}. For our trap geometry, a tuning ratio of $T_\mathrm{opt}=0.09$ minimizes the axial frequency dependence with respect to kinetic energy and therefore motional amplitudes around the center of the trap.  A complementary analysis shows that the first even-order coefficient of the anharmonicity $C_4$ (Ref. \cite{Ketter2014}) becomes zero near our calculated $T_\mathrm{opt}$.   

Ultimately, we elect to operate the science trap at a voltage of $V_\mathrm{STCR}=V_T=\SI{-65}{\volt}$ applied to the ring electrode and $V_\mathrm{STRW}=V_\mathrm{STCR}T_\mathrm{opt}=\SI{-5.85}{\volt}$ while all other electrodes are held at ground. This voltage choice is somewhat arbitrary but represents a compromise between trap depth and the desire to limit voltage swings during the shuttling procedure transporting ions from the loading trap to the science trap. The simulated and experimentally determined motional frequencies are listed in Table~\ref{Table:9BeSTfreq}.
Once ions are in the science trap we are able to operate with potential differences up to $\SI{800}{\volt}$ before we anticipate limiting leakage currents through the insulators, corresponding to a maximum axial frequency of \SI{1.4}{\mega\hertz}. Operation at \SI{260}{\volt} yields axial frequencies of about \SI{800}{\kilo\hertz} as used in previous studies \cite{Britton2012}, while maintaining magnetron frequencies below \SI{100}{\kilo\hertz}.  

\begin{table}[bp]
\begin{center}
\renewcommand{\arraystretch}{1.25}
\begin{tabular}{r r r}\hline
\multirow{2}{*}{Eigenmotion} & \multicolumn{2}{c}{Frequency (\SI{}{\kilo\hertz})} \\
 & Simulation & Experiment \\
 \hline
Axial $\nu_z$       & 406.4 & 402 \\
Magnetron $\nu_-$   & 24.4 & 23.9\\
Cyclotron $\nu_+$   & 3381.5 & 3382\\
\hline

\end{tabular}
\caption{Simulated and experimentally determined motional frequencies of a \IonBep{} ion in a magnetic field of $B_0=\SI{1.998}{\tesla}$ and at a science trap potential of $V_\mathrm{STCR}=\SI{-65}{\volt}$ and a tuning ratio of $T_\mathrm{opt}$. The calculated $C_2=\SI{-4.68e-3}{\milli\meter^{-2}}$ deviates from the measurement by $\approx\SI{2}{\percent}$. The magnetron and the reduced cyclotron frequency have been determined from the measured axial frequency and magnetic field. The free cyclotron frequency is $\nu_c\approx\SI{3406}{\kilo\hertz}$.   }
\label{Table:9BeSTfreq}
\end{center}

\end{table}

\subsection{Ion loading and shuttling}
Ion loading commences by biasing the loading-trap electrodes using low-noise, computer-controlled, high-voltage power supplies (iseg EHQ 102M) .  The loading trap center ring voltage is held at only a few volts (typically $V_\mathrm{LTCR} \geq$\,\SI{-10}{\volt}) while other loading trap electrodes are held at ground.  This produces a shallow electrostatic potential difference which restricts the kinetic energy distribution of ionized beryllium captured during the loading process. During loading, Science-trap end cap 2 can be negatively biased (typically \SI{-100}{\volt}) to repel any electrons produced during the loading procedure either from the ionization process or photoemission due to stray light impinging on trap electrodes. 

A heating current of $\leq\SI{1}{\ampere}$ is applied to the beryllium oven described in section~\ref{Sec:LoadingTrap}, heating the beryllium wire to $\approx\SI{920}{\degreeCelsius}$.  This produces a flux of neutral $^9$Be estimated to be of order \SI{1e14}{\second^{-1}\meter^{-2}} near the trap center, taking into account the geometry of the loading trap. In test setups we have operated the oven with currents up to $\approx\SI{1.5}{\ampere}$ before beryllium plating became visible by eye on a proximal window. 

Atomic fluorescence near \SI{235}{\nano\meter} is monitored via a photomultiplier tube aligned to the sideview imaging port of the loading trap, with the photoionization laser turned on. The axial Doppler cooling laser is turned off during the loading cycle to prevent heating of radial motional modes when ions are created. The integrated atomic fluorescence observed on the photomultiplier tube provides a proxy measure for the number of ions created in the trap. We obtain $\approx10-100$ ions in the science trap (accounting for ion losses induced in the shuttling procedure described below) after $\leq\SI{3}{\min}$ of ionization time with photoionization laser powers around \SIrange{1}{4}{\milli\watt}.  

\begin{figure}
		 \includegraphics[width = 8.5cm]{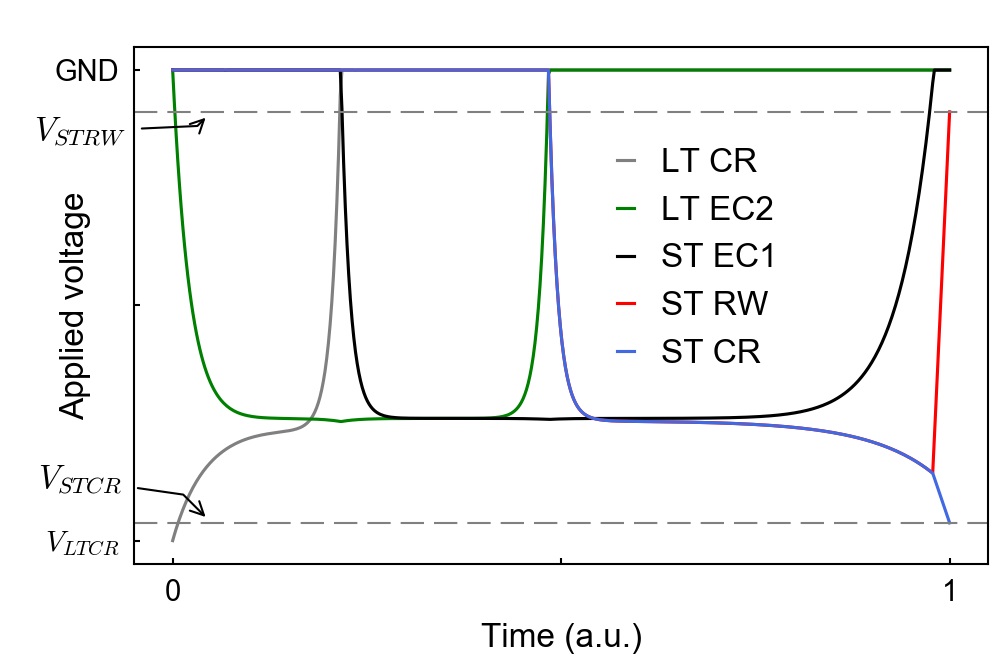}
    \caption{Potential sequence for quasi-adiabatic shuttling procedure simulations in SIMION. The science trap center ring and rotating wall electrodes are kept on the same potential throughout the sequence. Their final values, determined by the desired ion motional frequencies and the tuning ratio, are set at the end. $V_\mathrm{STCR}$ is the final center ring voltage, while $V_\mathrm{LTCR}$ is the loading trap voltage at the start of the shuttling procedure. To produce the same central potential in the loading trap as is required in the science trap, numerical values are given in Table~\ref{Table:potentials}. The potential depth as well as the total shuttling time are scalable. To achieve the \SI{-50}{\volt} moving potential referenced in the text we start with a loading trap center ring (LT CR) voltage $V_\mathrm{LTCR}=\SI{-67.55}{\volt}$ to produce this and finish with the science trap voltages of $V_\mathrm{STCR}=\SI{-65}{\volt}$ and $V_\mathrm{STRW}=\SI{-5.85}{\volt}$ as described above.  Electrodes not shown remain on ground throughout the procedure. LT: Loading trap, ST: Science trap, CR: Center ring, EC: End cap, RW: Rotating wall}
    \label{fig:Shuttling}
\end{figure}

Ion shuttling to the science trap proceeds as a sequential change in the voltages on all intermediate electrodes to create a moving potential well. The shape and temporal distribution of these potentials is simulated in SIMION such that the well moves at a constant velocity and depth along the trap axis.  The shuttling-potential sequence only determines the relative values for the potentials, which can be scaled to accommodate different ramping speeds. The slew rate with which the electrode potentials are changed should stay well below the motional frequencies of the ions to ensure adiabaticity. A schematic illustration of the sequence of shuttling voltages employed is shown in Fig.~\ref{fig:Shuttling}. In this work shuttling typically finishes after one minute with a moving potential depth throughout the procedure of \SI{-50}{\volt}. Table~\ref{Table:potentials} lists all voltages applied during ionization in the loading trap, shuttling from loading to science trap and trapping in the science trap.

\begin{table}[h]
\begin{center}
\renewcommand{\arraystretch}{1.25}
\begingroup
\setlength{\tabcolsep}{12pt} 
\begin{tabular}{l r r r}\hline
\multirow{2}{*}{Electrode} & \multicolumn{3}{c}{Voltage (V) during:} \\
 & loading & shuttling & trapping \\
\hline
LT EC1  & 0 & 0 & 0 \\
LT CR  & -10 & -67.55 & 0 \\
LT EC2  & 0 & -50 & 0 \\
ST EC1  & 0 & -50 & 0 \\
ST RW  & 0 & -57.86 & -5.85 \\
ST CR  & 0 & -57.86 & -65 \\
ST EC2  & 0 & 0 & 0 \\
\hline
\end{tabular}
\endgroup
\caption{Voltages applied to electrodes during ion loading, shuttling (maximum values given, for time dependence see Fig.~\ref{fig:Shuttling}) and trapping. LT: Loading trap, ST: Science trap, CR: Center ring, EC: End cap, RW: Rotating wall. 
}
\label{Table:potentials}
\end{center}

\end{table}

Upon completion of the automated shuttling sequence the ions will have moved into the path of the radial Doppler cooling beam in the science trap, at which time the axial cooling beam is turned on. Initial cooling proceeds by moving the radial laser beam from the edge of the trap towards the center while repeatedly sweeping the cooling-laser frequency from several GHz red-detuned towards the transition frequency.

Ionic fluorescence from resonant excitation can be monitored on a photomultiplier tube or camera on either sideview or topview imaging ports. After a few scanning cycles, ions occupying large magnetron radii converge on the trap center, and the cooling lasers are moved to the optimum Doppler cooling frequency: half a linewidth $\gamma/2\approx \SI{10}{\mega\hertz}$ below resonance. Adjusting laser frequencies during the cooling procedure allows the relevant atomic transition frequencies to be measured, as summarized in Table~\ref{Table:BeCalcs} in Appendix~\ref{Appendix:BeLevels} along with the calculated values.

The observed ion lifetime, which is dominantly limited by charge-exchange reactions with H$_2$ during Doppler cooling to the \tptt{} excited state \cite{Sawyer2015}, is consistent with our background gas pressure of $\sim\SI{6e-11}{\milli\bar}$. Under constant application of both Doppler cooling beams with total intensity of approximately $0.8\times$ the saturation intensity, we observe a mean single-ion lifetime of about 1.1 hours, extracted via observation of fluorescence with a photomultiplier tube connected to the science trap sideview imaging system. As centripetally-separated heavy ions are only sympathetically cooled, a buildup of contamination ultimately destabilizes the ion crystal over several hours of continuous cooling.  

\begin{figure*}
		\includegraphics[width=17cm]{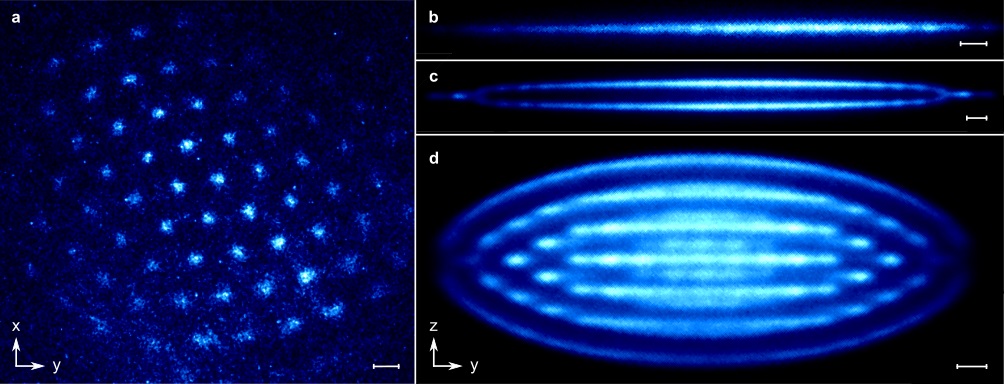}
    \caption{Images of \IonBep\ crystals taken with topview (a) and sideview (b - d) imaging systems. (a) Site-resolved ion crystal showing triangular lattice structure when camera is gated at the crystal rotation frequency. The crystal rotation frequency is locked to the rotating wall at about \SI{26.6}{\kilo\hertz} with \SI{100}{\nano\second} exposures per trigger over \SI{9}{\second}. (b-d) Single-plane, double-plane and multi-plane ion crystal conformations as seen on the sideview for several seconds of continuous exposure. Images (a) and (b) show the same crystal. Axes indicate orientation in space, size bars are \SI{20}{\micro\meter} long in all images.}
    \label{Fig:Crystals}
\end{figure*}

\subsection{Site-resolved imaging of ion crystals}

Ions crystallize within a few seconds of the completion of ion shuttling and application of the correctly aligned Doppler cooling beams.  Time-averaged sideview imaging may be performed directly; the sharp focus of the sideview imaging system restricts light collection to a narrow volume in space.  This results in a sectional view with individually resolvable ion planes, and indicates that ion trajectories are stable over many seconds, as required for light collection in Fig.~\ref{Fig:Crystals} via an Andor iXon Ultra 897 camera.

The crystal conformation, or ion density, is determined by the torque applied from cooling lasers, field imperfections, and the rotating wall potential, while the typical axial ion spacing is also a function of the trap potential. Crystals rotating at frequencies $\Omega$ close to the high or low frequency deconfinement edges will have lenticular shapes (see Appendix~\ref{Sec:IdealPenningTrap}, reducing to a single plane for sufficiently low rotation frequencies as in Fig.~\ref{Fig:Crystals}(b).  Beyond this low-frequency limit, ion-ion spacings diverge and ions are lost radially.  By contrast, crystals rotating at frequencies closer to the middle of the permissible range will have larger axial extent as in Fig.~\ref{Fig:Crystals}(d).

Continuous exposure during topview imaging suffers from rotational averaging of the ion locations about the trap axis.  To overcome this we synchronize a gateable camera (Andor iStar ICCD) to the ion rotation rate in order to image stroboscopically.  This requires precise control of the rate of rotation, beyond what is typically achievable using radial lasers alone \cite{mavadia2013}. We implement the rotating wall drive method as it gives a direct and precise rotation-frequency control, improved robustness to laser torque changes from intensity or frequency fluctuations,  a convenient source for camera synchronization, and additionally yields lower crystal temperatures when used in conjunction with a cooling laser \cite{Torrisi2016}. 

The camera is triggered at half the rotating wall drive frequency to match the crystal rotation frequency, as we use a quadrupole rotating wall.  The gating period determines the arc subtended by individual ions during the imaging period and should be kept short to permit single-ion resolution at the crystal periphery where velocities are highest.  These velocities in turn depend on the rotation frequency of the crystal, and hence the imaging trigger frequency, as the inter-ion distance also changes with rotation frequency.  A trade-off between improved angular resolution and reduced exposure time must be optimized for a particular crystal radius and applied trapping potential. 

An image of a locked, planar ion crystal is shown in Fig.~\ref{Fig:Crystals}(a) with about 70 ions in a triangular lattice of $\approx\SI{35}{\micro\meter}$ radial spacing, owing to a rotation frequency close to the edge of deconfinement.  After careful optical alignment of the radial cooling laser, we observe locked, stable crystals over tens of seconds. For Fig.~\ref{Fig:Crystals}(a) we restricted the camera gate width to \SI{100}{\nano\second} at a rotation frequency of about \SI{26.6}{\kilo\hertz}, yielding an angular resolution of $\approx\SI{1}{\degree}$. In these images the ion image size is well above the value predicted by the system design. Interferometric tests on the wavefront error of the inbore imaging sections suggests this is not due to the imaging system itself, which is supported by extrabore resolution testing using a 1951 USAF test target.  It is therefore likely that suboptimal imaging performance is due to other factors such as imperfect alignment relative to the trap, vibrations, elevated in-plane temperatures, or fast crystal fluctuations.

\subsection{Vibration measurements}

\begin{figure*}
\begin{center}
		\includegraphics[width=17cm]{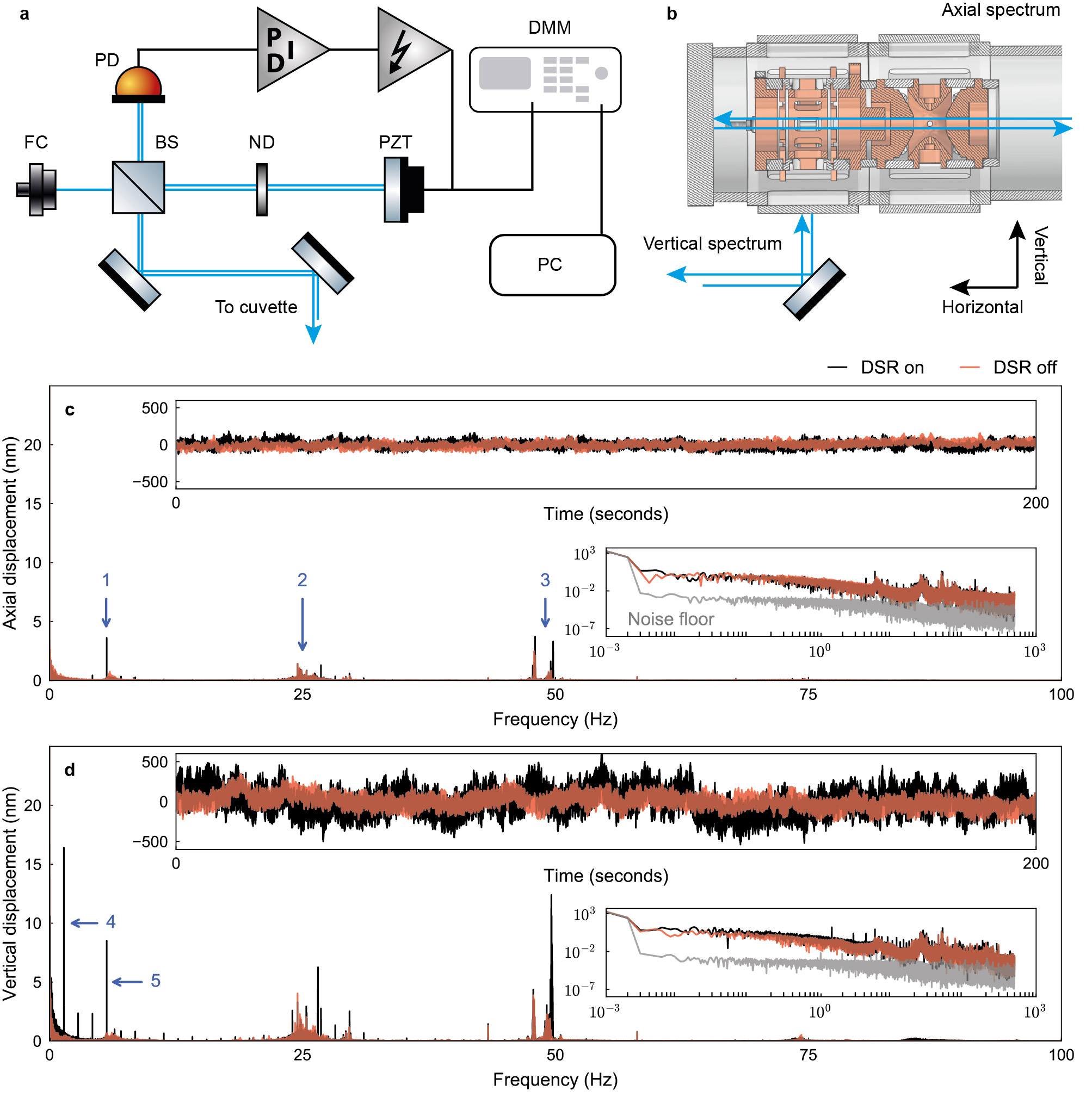}
    \caption{Interferometric measurement of differential motion between vacuum envelope and optomechanics. (a) Michelson interferometer setup. Light from a fiber coupler (FC) hits a beamsplitter (BS), where it is partially reflected towards the cuvette in the test arm. The transmitted portion is adjusted in power by a neutral density (ND) filter and reflected off a mirror mounted on a piezo actuator (PZT) in the reference arm. Both arms combine on the photodiode (PD) whose signal is fed to a feedback circuit, amplified and fed back to the actuator to stabilize the phase on the detector. The piezo signal is read through a digital multimeter (DMM) and send to the PC. Differential motion on the interferometer breadboard itself was determined with a fixed mirror in the measurement arm to be below \SI{1}{\nano\meter} between \SIrange{2}{950}{\hertz}, such that the spatial resolution in the measurements in this band is limited by the detection and digitization noise rather than motion in the reference arm. (b) Beam trajectories for axial and vertical vibration spectra. (c) Axial displacement spectrum with DSR turned on and off along with representative time traces. Data beyond \SI{500}{\hertz} measured using a different apparatus show no discrete features. Upper inset: first $\SI{20}{\percent}$  of the time domain data used to produce spectra in the main panel.  Drift-corrected root-mean-square (RMS) deviations are \SI{33}{\nano\meter} with DSR on and \SI{28}{\nano\meter} with DSR off, as we observe the slow drifts in data are due to thermal expansion of the trap mounting system.  Lower inset: logarithmic representation of main panel data with extended bandwidth. (d) Same as (c) for vertical displacement. Drift-corrected RMS deviations are \SI{117}{\nano\meter} with DSR on and \SI{56}{\nano\meter} with DSR off. All time traces were sampled at \SI{1}{\kilo\hertz} for \SI{1}{\kilo\second} without averaging. Indicated peaks show mechanical resonances of the support structure holding the DSR (1), acoustic noise from the cryogens (2), compressors and chillers for cryo-cooler cycles (3) in the service space outside the lab, mechanical motion from the pulse-tube (4) and its coupling to the mechanical resonances (5).}
    \label{fig:Vibrations}
\end{center}
\end{figure*}

Differential motion between the optomechanics and the ion crystal is a known challenge in quantum simulation experiments with ion crystals.  The resulting angular misalignment between the wavefronts of the Raman interference pattern used to engineer spin-spin couplings must be minimized on all relevant timescales.  Accordingly, we have characterized the vibration-induced motion of the glass cuvette relative to the optomechanical system in order to bound expected performance.  We pay particular attention to motion both in the vertical direction transverse to the trap axis (corresponding to the direction with maximum mechanical compliance), and along the trap axis - both of which could cause errors in alignment of Raman wavefronts relative to the ions.

We determine this differential motion by measuring the vibrational spectra of the glass cuvette as a proxy for trap motion via a Michelson interferometer setup as shown in Fig.~\ref{fig:Vibrations}. We obtain displacements signals via direct readout of the voltage applied to a calibrated piezo actuator in the reference arm of the interferometer.  The relative phase, and thus intensity on the photodiode, is locked via a digital feedback circuit on the piezo actuator, and the voltage signal is digitized and recorded via a digital multimeter (Keyisght 34465A) at \SI{1}{\kilo\hertz} for \SI{1}{\kilo\second}.

Measurements are taken consecutively in three different axes: along the trap axis (Fig.~\ref{fig:Vibrations}(c)), at a right angle in the vertical direction (Fig.~\ref{fig:Vibrations}(d)) and at \SI{45}{\degree} to the horizontal (same results as Fig.~\ref{fig:Vibrations}(d), not shown). The consecutive nature of the measurements means that reconstructing the vibrational spectrum in the horizontal is not possible via vector addition, but we observe quantitative agreement between measurements in the two transverse directions.  Comparative measurements are performed with the dual-stage reliquefier and associated infrastructure either in operation or turned off, in order to determine the contribution of vibrations from this cryogenic system.

Both axial and vertical measurements show root-mean-square deviations of tens of nanometers over hundreds of seconds (Fig.~\ref{fig:Vibrations}(c-d), upper insets).  Closer examination in the Fourier domain reveals that vibrations show a number of discrete spectral features on a vibrational baseline above the electronic background (scaling as $1/f$), in the \SIrange{1}{100}{\hertz} region as identified in Fig.~\ref{fig:Vibrations}.  The individual magnitudes of these spectral features stay below $\approx\SI{20}{\nano\meter}$ in the transverse direction and below $\approx\SI{5}{\nano\meter}$ in the axial. There are no discrete spectral features evident below \SI{1}{\hertz} and all spectral features above \SI{100}{\hertz} are less than \SI{1}{\nano\meter} in amplitude and hence not shown. Vertical displacement spectra are in good agreement with measurements conducted via accelerometer measurements on the laboratory floor and optical table.

The vertical displacement spectrum features prominent spectral features at the pulse-tube frequency ($\approx\SI{1.4}{\hertz}$) and its harmonics, which are present also as sidebands around various spectral features. Notably, these features are strongly suppressed in the axial direction, i.e. in the direction more critical for phase stability of the optical dipole force. Sharp spectral features corresponding to vibrations from electro-motors in both the compressor and chiller for the pulse-tube are seen around \SI{50}{\hertz} and can be strongly suppressed by switching them off. Broad spectral features around \SI{25}{\hertz} are believed to stem from the cryogen reservoir as these features persist in a power down of nearly all devices in the lab, but grow considerably if the magnet overpressure is vented during the DSR shutdown. Exacerbation of the harmonics of the pulse-tube around \SI{5.6}{\hertz} are suspected to stem from a mechanical resonance of the support structure on which the DSR rests.  Finally, we have found that careful damping of the high-pressure lines connecting the DSR head to the compressor has the most significant impact on measured vibrations transmitted to the trap.



\section{Conclusion and outlook}\label{Sec:Conclusion}  
In this manuscript we have reported on the design, construction, and characterization of the core elements of an experimental system for quantum simulation based on trapped-ion crystals in a Penning trap.  Our presentation focused on novel elements, such as the development of a co-designed high-optical-access trap and inbore optomechanical system targeting the laser configurations needed in quantum simulation.  In addition, we introduced a world-first dual-stage liquid-cryogen reliquefier enabling uninterrupted operation over many months.  There is no measurable degradation in magnet performance (e.g. homogeneity and temporal stability), to within limits imposed by our measurements of an NMR-signal linewidth. Moreover, we have performed detailed vibration measurements using  
an interferometric technique, complemented by accelerometer measurements, to demonstrate that the use of this cryogenic configuration does not substantially degrade trap stability, though this system also flexibly allows free-running operation without the DSR as may be needed for the most sensitive measurements.

We have demonstrated ion crystals as well as site-resolved imaging using a custom objective and phase-locking of the crystal rotation to a rotating-wall potential.  Image formation is stable over tens of seconds, which is promising for future experiments where both crystal and system mechanical stability are important.  Challenges remain in improving upon the demonstrated core capabilities, including detailed analysis of the impact of the residual measured vibrations on image quality and qubit coherence\cite{BrittonMagnet2016}, and optimizing alignment of the imaging system in order to improve detection signal-to-noise ratios.

The system reported here represents a work in progress and clearly does not constitute a functional quantum simulator at this stage. We identify four areas for future development where ongoing effort is required in order to bring this system to a practical level of functionality (these are not exhaustive): (i) Crystal stability through magnetic field alignment, (ii) Raman beam delivery and optical dipole force wavefront alignment, (iii) sub-Doppler cooling, and (iv) imaging of ion crystal dynamics.

\begin{figure}[h]
		\includegraphics[width = 8.5cm]{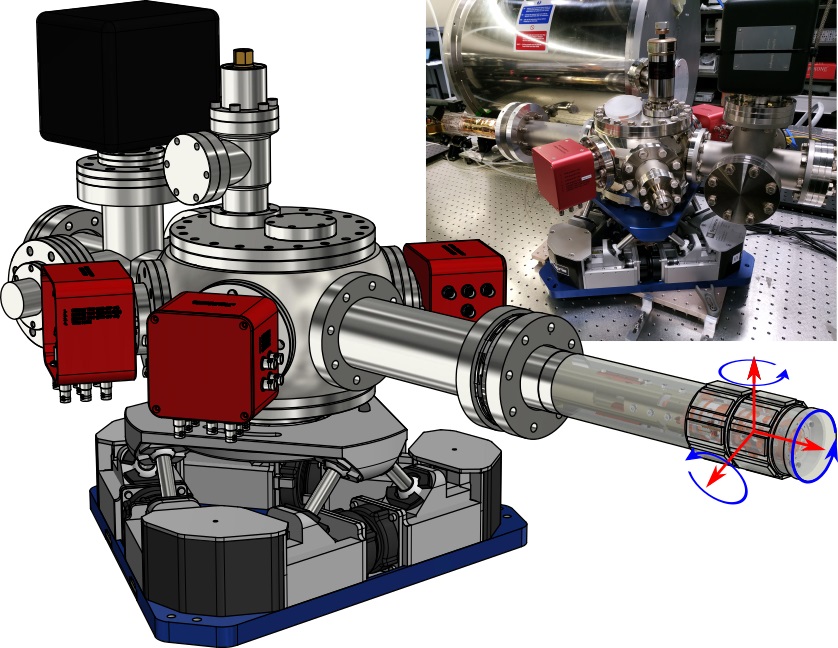}
    \caption{Model representation and photograph of ultra-high-vacuum chamber and trap tower on hexaglide six-axis precision positioning system, motional degrees of freedom indicated at ion location.}
    \label{fig:Hexaglide}
\end{figure}

\begin{enumerate}[i]
    \item Plasma heating and crystal instability are known to arise from misalignment of the electric and magnetic field axes in a Penning trap \cite{Bollinger1993,Huang1998}; target tolerance $\leq\SI{0.01}{\degree}$. In order to improve upon our existing manual alignment procedure we have now integrated a custom-engineered non-magnetic, absolute-encoding hexaglide system (see Fig.~\ref{fig:Hexaglide}) on which the trap ultra-high-vacuum system is mounted for precise positioning in all spatial degrees of freedom. Interferometric testing demonstrates $\leq\SI{1}{\micro\meter}$ translational and sub-millidegree rotational positioning resolution and repeatability at arbitrary pivot points.  We have implemented a limit-checking algorithm to prevent collisions of the trap, optomechanics, and magnet for arbitrary pivot points \cite{Tang2018}. 
    
    During the review of this manuscript, we have performed millimeter-wave-driven coherent qubit operations to more precisely map the magnetic field inhomogeniety experienced by the ions in their environment. Using a single-plane ion crystal translated electrostatically along the trap axis we find a linearized, relative magnetic field gradient in the axial direction of $\approx\SI{3.2e-7}{\per\milli\meter}$. We additionally find a crystal of $\approx\SI{300}{\micro\meter}$ diameter at millimeter-wave excitation times of $\approx\SI{2}{\milli\second}$ to still be Fourier-limited at $\approx\SI{0.5}{\kilo\hertz}$ in its resonance linewidth, providing an upper bound for radial field imhomogenieties. More detailed mapping of the spatial dependence in the radial plane may be possible using quasi-static cloud displacement through the rotating wall electrodes. Knowledge of these gradients and qubit control capabilities for sensing allow for in situ minimization of the magnetic field distortions using the existing shimming coils. The details of the millimeter-wave system will be described elsewhere~\cite{Marciniak2019PhD}.

    \item Raman beam alignment relative to the ion crystal and wavefront stability are important parameters for effective realization of the optical dipole force used to engineer Ising Hamiltonians.  Tolerances on angular alignment shrink as we exploit the enhanced optical access associated with wide opening angles $\theta_R$ through the decrease in effective lattice wavelength to an achievable minimum of $\lambda_{R}\approx\SI{560}{\nano\meter}$. Vibrational motion along the trap axis contributes to the question of whether we achieve the Lamb-Dicke confinement criterion in this setup.  Assuming an RMS motion defining the effective axial extent of an ion in the crystal, $z_{RMS,i}\approx \SI{30}{\nano\meter}$ (see Fig.~\ref{fig:Vibrations}), the individual ion Lamb-Dicke confinement parameter for $\theta_{R}=\SI{32}{\degree}$, $\eta_{i}=(2\pi z_{RMS,i}/\lambda_{R})\approx0.4$, slightly better than values achieved in~\cite{Britton2012} (there limited by thermal excitation).  Transverse vibrational motion can produce inhomogeous phase differences across the crystal due to either pivoting of the trap cantilever, or any finite mismatch angle.  Vibration-induced tilts of $\approx\SI{0.0001}{\degree\per\micro\meter}$ are small compared to the axial motion and also to the expected wavefront alignment error (equal or better than $\approx\SI{0.05}{\degree}$ from~\cite{Britton2012}). Nonetheless the impact of vibrations in this system must be carefully managed on experimentally relevant timescales (due to the discrete vibrational frequency content of the spectrum) and represent an opportunity for further improvement in the system.
    
    We plan to add deflector mirrors mounted on piezo-actuated goniometers equipped with resistive encoders for computer-controlled, precision inbore beam alignment with typical minimal step sizes down to \SI{0.0001}{\degree}. Inbore adjusters have the further advantage of reducing the throw into the bore, and reduce differential movement between mirrors and trap. This will enable automated beam-position searches when performing alignment, in situ beam adjustment, and computer-controlled orientation of the Raman wavefronts relative to the crystal~\cite{Britton2012}. 

    \item If our trap is operated at the measured axial frequency of $\sim \SI{400}{\kilo\hertz}$, and we assume mode temperatures similar to those reported in~\cite{Britton2012}, the thermally induced $z_{RMS,i}$ will increase by a factor of $\approx\sqrt{2}$.  Achieving $\eta_{i}<1$ then limits the useful value of $\theta_{R}$ to $\approx 4.5$ degrees, overwhelming other advantages.  We thus identify the implementation of new sub-Doppler cooling techniques as an important line of further inquiry for our system. Recent demonstrations of electromagnetically-induced-transparency cooling of the axial modes of large ion crystals\cite{EITtheory,EITexperiment} are well suited for our geometry and promise to allow for increasing Raman beam opening angles without encountering thermal limitations.  
    
    \item Studying spin dynamics will necessitate novel forms of high speed imaging.  Single-shot imaging at frame rates comparable to or exceeding the crystal rotation frequency are coming within reach given advances in detector technology. Single photon avalanche photodiode (SPAD) arrays are an active research area, and early commercial offerings are now available.  We have obtained a SPAD array (SPC3, Micro Photon Devices) which comprises a \mbox{64 $\times$ 32} pixel array with \SI{96}{\kilo\hertz} frame rate and a quantum efficiency of $\approx\SI{20}{\percent}$ at \SI{313}{\nano\meter}.  Technical limitations, including a low fill factor of only \SI{4}{\percent} make these devices challenging to implement.  We are exploring integration of a microlens array (and appropriately designed imaging optics) to increase the effective fill factor, and potentially enable single-photon counting capabilities.  Further advances in this technology are likely to be of great interest for the pursuit of quantum simulation.

\end{enumerate}

 We are excited for future developments and hope that the results reported in this manuscript will enable other teams interested in Penning trap physics to enter the field. 

\begin{acknowledgments}
The authors acknowledge the assistance of Torsten Gaebel and David Reilly with NMR measurements and provision of the relevant equipment, Sandeep Mavadia for discussions, and John J. Bollinger for technical advice and input into system design and operation. The project was supported by the ARC Centre of Excellence for Engineered Quantum Systems (CE110001013) and a private grant from H. and A. Harley.
\end{acknowledgments}



\bibliography{references}

%
%
\begin{widetext}
\appendix
\newpage

%

\section{Beryllium energy-level-structure calculation}\label{Appendix:BeLevels}

The energetic structure of ion crystals in a Penning trap is intimately tied to the trap geometry and present potentials, which mandates simulations to make quantitative predictions of mode frequencies. However, the field-free transition frequencies in neutral and singly charged $^9\mathrm{Be}$ are well known \cite{Kramida1999} and magnetic-field-induced corrections to the level structure can be readily calculated due to the comparatively simple Hamiltonian for low mass number alkali-earth atoms and ions. Only $S$ and $P$ orbitals need to be considered in these calculations since all relevant transitions take place in the ground- and first-excited-state manifolds. In order to find the required frequencies for the resonant light-ion interactions, we need to find the eigenenergies of the ionic Hamiltonian for the \tsot\ and the \tptt\ states. This Hamiltonian is given by:
\begin{align}
	\HH' &= \HH_0+\HH\\
	\HH &= \HH_{SO}+\HH_{Ze}+\HH_{Zn}+\HH_{HF}
	\label{Hamiltonian}
\end{align}
where $\HH_0$ is the field-free Hamiltonian whose eigenenergies are known, and $\HH$ is the Hamiltonian that describes the perturbation by the magnetic field. It is has four contributions: $\HH_{SO}\propto \mathbf{L}\cdot\mathbf{S}$ is the electronic spin-orbit coupling term, $\HH_{Ze}=\mu_B(g_L\mathbf{L}+g_S\mathbf{S})\cdot\mathbf{B}$ is the Zeeman interaction of the electron's spin and orbital magnetic moments with the external magnetic field, $\HH_{Zn}=\mu_Bg_I\mathbf{I}\cdot\mathbf{B}$ is the Zeeman interaction of the nuclear magnetic moment with an external magnetic field, and $\HH_{HF}\propto \mathbf{I}\cdot\mathbf{J}$ is the hyperfine interaction. $h\mathbf{L}, h\mathbf{S},$ and $h\mathbf{I}$ are the orbital, spin and nuclear angular momentum operator, respectively, $h$ is Planck's constant, $\mu_B$ is the Bohr magneton, and the $g$'s denote associated $g$-factors of angular momenta. In the following we will briefly outline the derivation of transition frequencies.

In our intermediate field strengths, where neither $\HH_{SO}\gg\HH_{Ze}$ nor $\HH_{SO}\ll\HH_{Ze}$, we find the excited (\tptt{}) state energies by exact diagonalization of the electronic part of $\HH$ in the $\ket{L,S,m_L,m_S}\equiv\ket{m_L,m_S}$ basis, and treat the contributions arising due to nuclear magnetic moments as first order perturbations afterwards. This is a good approximation because the energy scales in these two Hamiltonians differ by several orders of magnitude for the \tptt{} state. The pertinent energies $E_{m_J}^{J}$ from this are those whose $m_J$ are $3/2$ for Doppler cooling and readout, and $1/2$ for the repump transition. They are given by
\begin{align}
E_{3/2}^{3/2} &= \tfrac{1}{3}\Delta E+\mu_B(1+\tfrac{1}{2}g_S)B \\
E_{1/2}^{3/2} &= \tfrac{1}{6}\left(-\Delta E+3\mu_BB+\Big[9 \Delta E^2\right.+\left. 6\mu_B\Delta E(g_S-1)B+9\mu_B^2(g_S-1)^2B^2\Big]^\frac{1}{2}\right)
\label{EP32}
\end{align}
where $\tfrac{1}{3}\Delta E$ is the constant of proportionality in $\HH_{SO}$. The nuclear magnetic moment perturbation introduces additional additive terms
\begin{equation}
	E_{m_I}^{m_J} = hA m_I m_J+\mu_B g_I m_I B,
	\label{eq:NuclearPerturbation}
\end{equation}
where $A$ is the hyperfine constant of proportionality in $\HH_{HF}$.

We cannot follow the same treatment for the ground state (\tsot{}) however, because the zero orbital angular momentum means the spin-orbit term $\HH_{SO}$ vanishes, therefore no longer dominating the energy scale. To find the ground state energies we consequently perform exact diagonalization arriving at a Breit-Rabi equation. The full Hamiltonian for the ground state is
\begin{align}
\HH &=hA\mathbf{I}\cdot\mathbf{J}+\mu_B(g_J\mathbf{J}+g_I\mathbf{I})\cdot\mathbf{B}\\
 &=hA\left[\frac{1}{2}(I_-J_++I_+J_-)+I_zJ_z\right]+\mu_B(g_JJ_z+g_II_z)B,
\label{S_Hamiltonian}
\end{align}
where in the last step we have rewritten the Hamiltonian in terms of ladder operators. From this representation it is clear that the stretched states with $m_F=\pm F$ are eigenstates in the $\ket{F,m_F}$ basis for all magnetic field strengths. To find the other energies we chose the $\ket{\pm}\equiv\ket{m_J=\pm\tfrac{1}{2},m_I=m_F\mp\tfrac{1}{2}}$ basis to conserve $m_F$ (as $\mathbf{F}$ commutes with $\HH$ for all magnetic field strengths) and diagonalize the matrix representation. The resulting energies are

\begin{align}
E_{F,m_F} =& -\frac{hA}{4}+g_Im_F\mu_BB-\frac{1}{2}\Big[ h^2A^2F^2+2hAm_F\mu_B(g_J-g_I)B+\mu_B^2(g_I-g_J)^2B^2\Big]^\frac{1}{2}.
\end{align}

Using the same approach we can also find the resonant step transition energies required for photoionization. The Hamiltonian is significantly simpler because neutral $^9$Be has electrons with anti-parallel spins in both ground \osz and excited \opo state such that the total spin $S = 0$ and therefore $\HH_{SO}=0$. Likewise $J=0$ in the ground state, such that the hyperfine interaction $A \mathbf{I}\cdot\mathbf{J}$ vanishes. The \opo hyperfine interaction can also be neglected since for a singlet state it can be shown quite generally that the hyperfine $A(^1L_J) = 0$ \cite{atomicstructure}. This yields the simple Hamiltonian
\begin{align}
\HH =& \mu_B\mathbf{L}\cdot\mathbf{B}+\mu_Bg_I\mathbf{I}\cdot\mathbf{B}\\
=& \mu_BL_zB+\mu_Bg_II_zB
\label{Eq:BeAtomHamiltonian}
\end{align}
with eigenenergies
\begin{align}
E=\mu_Bm_LB+\mu_Bg_Im_IB\text{.}
\label{atomenergy}
\end{align}

\begin{table}[t]
\begin{center}
\renewcommand{\arraystretch}{1.25}
\begingroup
\setlength{\tabcolsep}{12pt} 
    \begin{tabular}{c | c c c}\hline
    Transition & $\nu_\text{calc.}$ & $\nu_\text{meas.}$ & $\Delta\nu$ \\ \hline

    Photoionization & \SI{1276.052139(14)}{\tera\hertz} & \SI{1276.05209(1)}{\tera\hertz} & \SI{49}{\mega\hertz}\\

    Cooling & \SI{957.42523(12)}{\tera\hertz} & \SI{957.425014(20)}{\tera\hertz} & \SI{216}{\mega\hertz}\\
    
    Repump & \SI{957.44383(14)}{\tera\hertz} & - & -\\
    
    Qubit & \SI{55.062(28)}{\giga\hertz} & - & -\\
\hline
\end{tabular}
\endgroup
\setlength{\tabcolsep}{20pt}
\caption{Calculated and measured transition frequencies for atomic and ionic beryllium in a \SI{1.998}{T} magnetic field. The photoionization transition excites to the $m_L = -1$ state before exciting to the continuum.}
\label{Table:BeCalcs}
\end{center}
\end{table}

The calculated uncertainties in neutral beryllium are dominated by the magnetic field contribution of $\Delta B=\pm\SI{1}{\milli\tesla}$ for the $\Delta m_L = \pm 1$ transitions and the uncertainty in the field-free line of \SI{2.8}{\mega\hertz}\cite{WillWilliams} for the $\Delta m_L = 0$ transition. Uncertainties in ionic beryllium are dominated by the uncertainties of the literature value of the field-free transitions (90\% and 80\% for the cooling and repump transitions, respectively), and the magnetic field strength uncertainty. The qubit transition uncertainty is almost entirely due to the uncertainty in magnetic field strength.

The center of the atomic transition is experimentally determined by monitoring the maximum of the atomic fluorescence for a given flux and laser power during the loading process. Likewise, the center of the cooling transition is determined by monitoring the maximum of fluorescence peaks during the cooling procedure either in the loading trap or science trap. 

Laser frequency measurements are taken with a HighFinesse WSU-10 wavemeter measuring at the respective fundamental wavelengths around \SI{626}{\nano\meter} and \SI{470}{\nano\meter}. The wavemeter is periodically calibrated using a frequency stabilized Helium-Neon laser from SIOS Messtechnik GmbH with standard deviation $\le\SI{500}{\kilo\hertz}$. The experimental uncertainty is dominated by the wavemeter measurement uncertainty at the fundamental frequency.
Owing to the thermal creation process of the atomic flux, Doppler shifts will have a strong influence on the center of the photoionization line. The photoionization laser nominally strikes the flux at right angle, but for atoms of $\approx$~\SI{1000}{\kelvin} even sub-degree deviations from normal incidence will provide Doppler shifts of order of the linewidth. Our measured value for the photoionization transition deviates from the calculated normal incidence (or rest) transition frequency in neutral $^9$Be by $\approx$~\SI{49}{\mega\hertz}, i.\,e. less than one linewidth, which is consistent with Doppler shifts caused by angular misalignment of the center of the atomic emission cone by less than \ang{1}.

\section{Penning trap fundamentals}
\label{Sec:IdealPenningTrap}

In a Penning trap charged-particle confinement is achieved by the spatial superposition of an electrostatic quadrupole field $\Evec=-\grad\Phi$ providing confinement along a trap axis, oriented parallel to a static, homogeneous magnetic field $\Bvec=B_0\zhat$ which provides radial confinement in the \plane{xy} via the Lorentz force. This trap therefore produces stable three-dimensional charged-particle confinement.

The archetypical version of the Penning trap employs only an axisymmetric center ring electrode and two axisymmetric end cap electrodes with hyperbolic surfaces to define a quadratic three-dimensional potential. In practice additional correction electrodes are inserted between the center ring and end cap electrodes to compensate for field imperfections. The end caps and center ring are biased with respect to each other by application of an external voltage $V_T$, resulting in an electric potential $\Phi_{T,i}$ at the position of ion $i$ whose functional form is
\begin{equation}\label{Eq:IdealHarmonicTrappingPotential}
	\Phi_{T,i}(x_i,y_i,z_i) = V_T C_2 \left( z_i^2 - \frac{x_i^2 + y_i^2}{2}  \right),
\end{equation}
where the parameter $C_2$ derives from the trap geometry. The equation of motion of a single particle in a Penning trap results in three fundamental motional modes\cite{Major}: an axial mode oriented along the magnetic field with axial frequency
\begin{equation}
	\label{eq:axialfreq}
	\nu_z=\frac{\omega_z}{2\pi} = \frac{1}{2\pi} \sqrt{2 V_T C_2 \frac{q}{m}},
\end{equation}
where $q/m$ is the ion's charge-to-mass ratio, and two radial modes with eigenfrequencies
\begin{equation}
	\nu_\pm = \frac{\omega_\pm}{2 \pi} = \frac{1}{2} \left( \nu_c \pm \sqrt{\nu_c^2 - 2\nu_z^2} \right).
\end{equation}
Here \mbox{$\nu_-=\omega_-/2\pi$} is the magnetron frequency, \mbox{$\nu_+=\omega_+/2\pi$} is the reduced cyclotron frequency and \mbox{$\nu_c=\omega_c/2\pi=B_0q/2\pi m$} is the cyclotron frequency. Stable trapping is achieved as long as $qV_TC_2>0$ and $\nu_z<\nu_c/\sqrt{2}$.\\

An additional potential is in general applied in order to provide an electrical means to tune the ratio of radial-to-axial confinement without modifying either the quantizing magnetic field or the axial potential.  In the absence of any external excitations, total angular momentum is conserved and the ion cloud can reach a stable equilibrium with rotation frequency $\Omega$ anywhere between the single-particle magnetron and modified cyclotron frequencies, $\omega_{-}\le\Omega\le\omega_{+}$. The value taken by $\Omega$ depends on the initial conditions and the ion cloud density. 

Within this range the rotation frequency can be precisely locked using a so-called rotating wall potential~\cite{Huang1997,FreericksPRA2013}. The generation of such a potential, $\Phi_{W}$, in the radial plane involves applying phase-shifted radio-frequency signals. This effectively distorts the radial trapping potential and imposes an external torque forcing a steady-state rotation at the frequency of the applied signal~\cite{Huang1997,Hasegawa2005}. 

It is advantageous to implement the rotating wall as a rotating azimuthal quadrupolar electric field, whose potential in the \plane{xy} is given by
\begin{equation}
	\hspace*{-3mm} \Phi_{W,i} \left(x_i,y_i,t \right) = V_{W} C_{W} \left( x_i^2 + y_i^2 \right) \cos \left[ 2 \left( \theta_i + \omega_{W} t \right) \right].
\end{equation}
In this equation $V_{W}$ is the amplitude of the sinusoidal signal applied to the electrodes, $C_{W}$ is an electrode-geometry-dependent scaling factor, $\theta_i$ the azimuthal angle at the position of the ion and \mbox{$\omega_{W}=2\pi\nu_{W}$} is the angular frequency of the rotating wall. The combined potential at the position of the $i$th ion is therefore
\begin{equation}
	\Phi_i \left(x_i,y_i,z_i,t \right) =\Phi_{T,i} + \Phi_{W,i} + \Phi_{C,i}.
	\label{eq:PotPenning}
\end{equation}
The final term above accounts for the Coulomb interaction between all ions
\begin{equation}
	\Phi_{C,i} \left(x_i,y_i,z_i \right) = \frac{q}{8\pi\epsilon_0}\sum_{i\neq k}{\frac{1}{r_{i,k}}},
\end{equation}
with $r_{i,k}$ denoting the distance between ion $i$ and $k$.\\ 

We can remove the explicit time dependence of the potential by transforming  into a frame of reference rotating at $\omega_W$, which yields
\begin{equation}
\label{eq:PotRotPenning}	
\Phi_i^R  \left(x_i^R,y_i^R,z_i^R \right) = \frac{m}{2q} \bigg[ \omega_z^2 \left(z_i^R\right)^2 + \left(\omega_{eff}^2 + \omega_{WA}^2\right) \left(x_i^R\right)^2  + \left(\omega_{eff}^2 - \omega_{WA}^2\right) \left(y_i^R\right)^2 \bigg] + \frac{q}{8\pi\epsilon_0}\sum_{i\neq k}{\frac{1}{r^R_{i,k}}}.
\end{equation}
Here we have introduced the effective trapping frequency 
\begin{equation}
	\omega_{eff} = \sqrt{\omega_{W} \omega_c - \omega_{W}^2 - \omega_z^2 / 2},
\end{equation}
and a frequency scale associated with the rotating wall amplitude,
\begin{equation}
	\omega_{WA} = \sqrt{2 V_{W} C_{W} q/m}
\end{equation}
which results in ellipticity of the ion cloud. The presence of the rotating wall introduces the additional criterion $\omega_{WA} \leq \omega_{eff}$ that must be met for stable trapping. This condition may be equivalently rewritten in terms of a deconfinement frequency
\begin{equation}
	\omega_{dc} = \frac{1}{2} \left( \omega_c - \sqrt{\omega_c^2 - 2\omega_z^2 - 4\omega_{WA}^2} \right),
\end{equation}
where for ion-cloud rotation frequencies below this value radial confinement is insufficient to maintain stable trapping. 

Locking the global rotation frequency provides advantages in trapping as it prevents the plasma from spinning down under the ambient drag from static field errors and collisions with background neutrals~\cite{Huang1998}. In addition, when moving towards the realization of laser-cooled Coulomb crystals, stabilizing the ion rotation frequency enables spatially resolved crystal images, for example via stroboscopic measurements~\cite{ItanoScience1998} or a transformation in the co-rotating frame~\cite{Britton2012}. 

The trap-electrode design significantly departs from the ideal hyperbolic electrode geometries, inevitably distorting the ideal quadrupole potential.  We characterize anharmonic contributions to the potential using an expansion in spherical harmonics about the trap center
\begin{align}
\label{Eq:MultipoleExpansion}
\Phi = 
\sum_{n=0}^{\infty}\Phi_n=V_T\sum_{n=0}^\infty C_n\rho^nP_n\left(\frac{z}{\rho}\right),
\end{align} 
where $\rho = \sqrt{x^2+y^2+z^2}=\sqrt{r^2+z^2}$,  $z/\rho = \cos\Xi$ in spherical coordinates and $P_n(\cos\Xi)$ are the Legendre polynomials of order $n$. Assuming rotational symmetry, odd-ordered coefficients $C_n$ vanish, so that the first four non-zero terms take the form

\begin{align}
\label{Eq:C0Term}\Phi_0 & = V_T C_0\\
\label{Eq:C2Term}\Phi_2 & = V_T C_2\left(z^2-\frac{r^2}{2}\right)\\
\label{Eq:C4Term}\Phi_4 & = V_T C_4\left(z^4 - 3r^2z^2 + \frac{3r^4}{8}\right)\\
\label{Eq:C6Term}\Phi_6 & = V_T C_6\left(z^6 - \frac{15r^2z^4}{2} + \frac{45r^4z^2}{8} - \frac{5r^6}{16}\right).
\end{align}

\noindent Nonzero terms for $n>2$ capture higher order perturbations on the ideal quadrupole term $\Phi_2$, while the zeroth order term $\Phi_0$ captures a static offset associated with the applied voltage. 

The specific dimensions of our trap design were selected by a process of optimization on the electrode geometry, minimizing anharmonicites $C_{4}$ and $C_{6}$ with terms beyond eighth order ignored.  In practice (see Section~\ref{Sec:ScienceTrap}) this involved performing a variational search over any free parameters, \eg~trap dimensions, electrode separation distances or compensation voltage, while maintaining the design constraints on aperture sizes and additional electrode structures. For a given choice of design parameters, the trapping potential was simulated and the result fitted to Eq.~\ref{Eq:MultipoleExpansion}, truncated to the desired order. The values of the expansion coefficients $C_n$ for $n>2$ extracted from this fit were then iteratively minimized, with the results shown in Table~\ref{tab:MultipoleExpansionCoefficients}. 

\begin{table}[h]
\begin{center}
\begin{tabularx}{5cm}{Y@{\hspace{0.5cm}}rY@{\hspace{0.5cm}} l@{\hspace{1cm}}}
\toprule
\hline
$C_0$ &  0.78133 &\\
$C_2$ & \SI{-4.68e-3}{} &\SI{}{\milli\meter^{-2}}\\
$C_4$ & \SI{-1.58e-7}{} &\SI{}{\milli\meter^{-4}}\\
$C_6$ & \SI{1.10e-7}{} &\SI{}{\milli\meter^{-6}}\\
\hline
\bottomrule

\end{tabularx}
\caption{Multipole expansion coefficients for the potential in a $\Delta z=\pm\SI{5}{\milli\meter}$ region about the center of the science trap.}
\label{tab:MultipoleExpansionCoefficients}
\end{center}
\end{table}

\section{Ion crystal equilibrium configuration}\label{Appendix:Crystal}

In the crystalline regime (where the Coulomb coupling parameter $\Gamma > 172$) the equilibrium ion positions are well approximated by solving the Lagrangian of the system assuming zero temperature~\cite{FreericksPRA2013,FreericksPRA2015,Britton2012}. Let $\rvec_j$ denote the position vector of the $i$th ion written $\rvec_i =(x_i,y_i,z_i)$ in Cartesian coordinates. The Lagrangian for a system of $N$ ions in the laboratory reference frame takes the form 
\begin{align}
	\mathcal{L} = \sum_{i=1}^N
	\left[
	\frac{1}{2}m\dot{\rvec}^2_i
	-\qch\left(\Phi_i(\rvec_i,t)
	-\Avec\cdot\dot{\rvec}_i\right)
	\right]
\end{align}
where $\Phi_i(\rvec_i,t)$ is the total scalar potential experienced by the $i$th ion (see to Eq.~\ref{eq:PotPenning}). This can be solved as an equilibrium problem by transforming the Lagrangian to the rotating frame, yielding
\begin{equation}
	\mathcal{L}^{{R}} = \sum_{i=1}^N
	\left[
	\frac{1}{2}m\abs{\dot{\rvec}_i^{{R}}}^2-\frac{\qch B_\text{eff}}{2}
	\left(
	\dot{x}_i^{{R}}y_{i}^{{R}}-\dot{y}_{i}^{{R}}x_{i}^{{R}}
	\right)
	-\qch\Phi_i^{{R}}
	\right].
\label{Eq:CrystalLagrangianRotatingFrame}    
\end{equation}
Here $B_\mathrm{eff} = B_0-2\omega_{W}m/\qch$ is the effective magnetic field in the rotating frame, which depends on rotational angular frequency due to the Coriolis force for ions moving in the radial plane. The scalar potential in the rotating frame is given by Eq.~\ref{eq:PotRotPenning}. The equilibrium configuration of the ions is then determined by finding the local minima of the effective potential energy, namely the positions $\rvec^{{R}}_{0,i}$ corresponding to the minimum of the classical action $S = \int dt\mathcal{L}^R$. Numerical solutions can be found which, for a 2D crystal, converge to a regular triangular lattice. Ion equilibrium positions deviate from a perfect triangular lattice near the crystal periphery, and have an overall ellipticity due to $V_W$ \cite{FreericksPRA2013,FreericksPRA2015}, consistent with experimental observations\cite{Britton2012}. 

\section{UHV parts list}
\label{Appendix:UHVPartsList}
\noindent Here we provide the detailed parts list associated with Fig.~\ref{Fig:exploded_vacuum_assembly}.

\begin{table}[H]
\footnotesize

\begin{tabularx}{15.5cm}{@{} lY @{} lY @{}}

\toprule
\hline

1 	&  	 hexagonal vacuum chamber	 &	Kimball Physics Inc.  MCF800-SphHex-G2E6\\ 
2 	&  	 CF160/40 adaptor, $2\times$CF40 flanges	 &  	 Vacom GmbH SPE-10058267-10\\ 
3 	&  	 CF160/63 adaptor	 &  	 Vacom GmbH ZL160063-316LNS\\ 
4 	&  	 all-metal valve		 &  	 MDC 314004\\ 
5 	&  	 feedthrough -- 25-pin SUB-D	 &  	 Vacom GmbH SPE-CF63L-SUBD-25-DE-CE-CBG-316L\\ 
6 	&  	 5-way cross		 &  	 Vacom GmbH FWX63R-316LNS\\ 
7 	&  	 ion getter pump	 &  	 Varian/Agilent VacIon Plus 20 StarCell 9191145\\ 
8 	&  	 Bayard-Alpert ion gauge	 &  	 Granville Phillips 274042\\ 
9 	&  	 vacuum viewport	 &  	 Torr Scientific Ltd. 16121-1\\ 
10 	&  	 trap tower assembly	 &  	 Fig.~\ref{Fig:trap_tube_assembly}\\ 
11 	&  	 CF63 through flange &  	 Vacom GmbH MCF450-PWF/GG \\ 
12 	&  	 non-evaporative getter (NEG)	 &  	 SAES Capacitorr D400-2\\ 
13,14,15 	&  	 feedthrough  -- 7 pin/12 kV/13 A	 &  	 Vacom GmbH CF40-MPCHV12- 7-SE-CE-MO\\

16,17,18\hspace{0.5cm} 	&  	SHV/BNC breakout boxes	 &  	custom\\ 

19 \& 20 	&  	CF63 flange \& glass cuvette	 &  Precision Glassblowing Inc.	 \\ 

\hline
\bottomrule
\end{tabularx}
\caption{Vacuum system part index for Fig.~\ref{Fig:exploded_vacuum_assembly}.}
\label{Table:vacuum_system_parts_list}
\end{table}

\section{Imaging system technical design}
\label{Sec:AppendixImaging}

The performance of both imaging systems described in Section~\ref{Sec:Imaging} can be characterized by the modulation transfer function, the Strehl ratio, and the on-axis point spread function.  For well-corrected optical systems all three quantities scale with the wavefront error, but can be used to highlight different aspects of imaging performance.

\paragraph{Modulation transfer function:} The modulation transfer function is a measure of image modulation (also called contrast). The image of a harmonic variation in intensity for a modulation transfer function of unity will be perfectly resolved with no smearing, whereas a vanishing modulation transfer function means aberration smears the image to be uniformly grey. A modulation transfer function of 0.5 or above is commonly consider sharp, while the lower end is highly subjective, but $\approx$~2-6\% is often still considered resolvable by eye. The modulation transfer function is commonly plotted as a function of spatial frequency which may be used to gauge the loss in contrast for finer features or sharper edges. The highest attainable spatial frequency is a direct measure of smallest spatial resolution. The overall shape of the ideal, non-obstructed modulation transfer function is governed by the aperture stop (aperture) only, as it is the modulus of the Fourier transform of the point-spread function. For real imaging systems the modulation transfer function becomes a function of field position and plane of incidence, i.e. both sagittal or tangential planes.

\paragraph{Strehl ratio:} The Strehl ratio is a convenient measure for well-corrected objectives to quantify closeness to a system limited only by diffraction since it combines all effects into a single number. Conventionally a value of 0.8 is considered diffraction-limited as most observers will have difficulty distinguishing improved image quality beyond that point. The Strehl ratio is often given as a function of field coordinate to extract the field of view within which it remains above a threshold.

\begin{figure}[h]%
\begin{center}
\includegraphics[width=17cm]{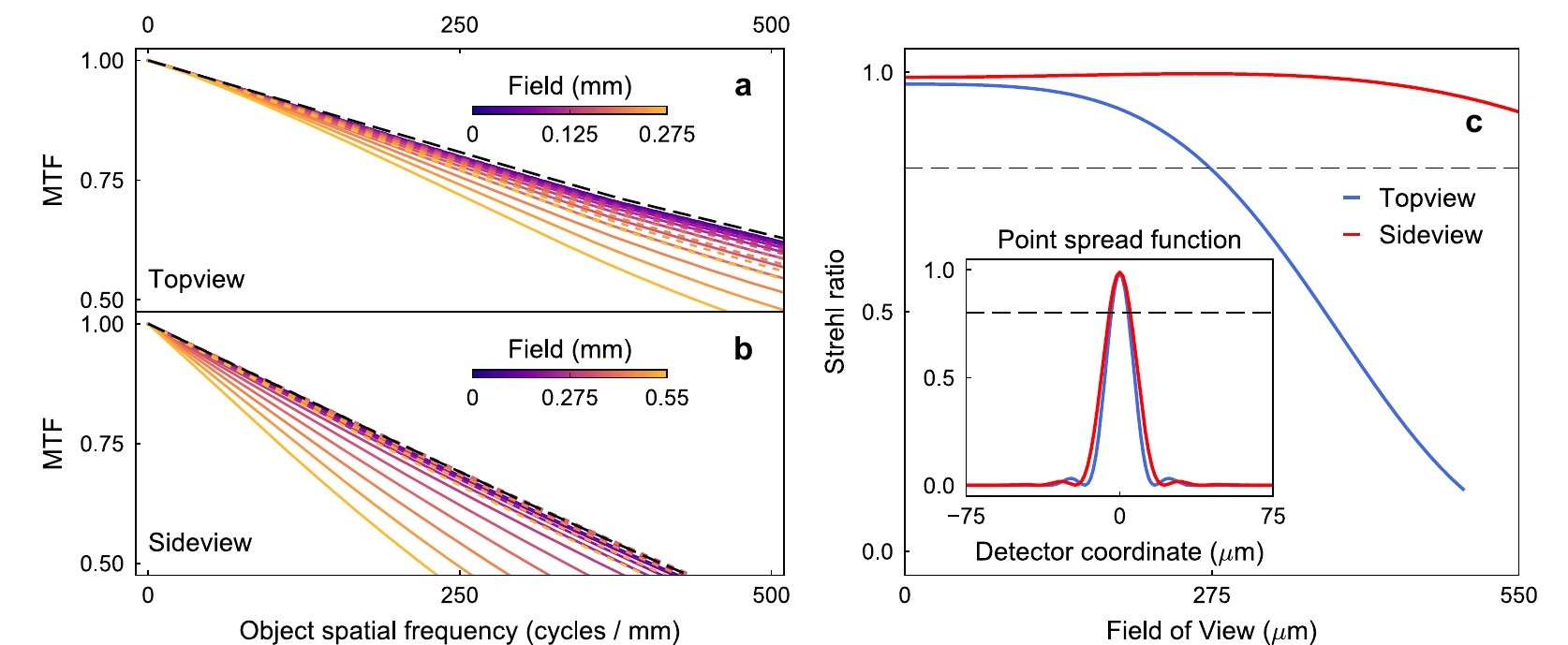}
\caption{(a,b) Modulation transfer function (MTF) of topview and sideview imaging as a function of object side spatial frequency. Solid lines are in the tangential plane, dashed lines in the sagittal plane. The black dashed line is the diffraction limited performance in both planes. (c) Strehl ratio for topview and sideview imaging systems as a function of field coordinate. Dashed black line corresponds to a ratio of 0.8 which is often used to define diffraction-limited performance. Inset: on-axis point spread function for topview and sideview as a function of detector coordinate. The peak value gives the Strehl ratio where diffraction limitation is indicated by the black dashed line. Note that the object side point spread function will be narrower by the magnification of the imaging systems.}
\label{Fig:ImagingFunctions}
\end{center}
\end{figure}

The design modulation transfer functions for both topview and sideview imaging systems are presented in Fig.~\ref{Fig:ImagingFunctions}(a)-(b), where we focus only on the high contrast region above 0.5. Both follow very closely the ideal shape for a circular field stop. The topview imaging has additional structure for mid-range spatial frequencies around 1200~cycles per mm from the presence of the deflection prism. The stronger dependence on field position and incidence plane in the sideview stems mainly from clipping on the center ring electrode before the imaging system. Note that the modulation transfer function is given in object space, i.e. demagnified, to give a direct indication of resolving power. For the systems presented here the spatial frequency at which the contrast falls to 0 is about 2000~cycles per mm for the topview, and 500 to 1000~cycles per mm depending on field point for the sideview, respectively.

We use the Strehl ratio to quantify over what field of view diffraction-limited performance can be expected. A plot of this is presented in Fig.~\ref{Fig:ImagingFunctions}(c) for both the topview and sideview systems, where the dashed line represents diffraction limited, with an inset that shows the on-axis point spread function for both systems in detector coordinates, i.e. scaled by system magnification. We note that the Strehl ratio gives information only about the relative performance to an aberration-free system of otherwise identical construction. The consistently high Strehl ratio for large field values in the sideview imaging indicates that performance can not be much improved.  However, the observed decrease in spatial cut-off frequency for large field points in the modulation transfer function plot shows that the expected image quality will fall towards the outer edges of the field of view.

Inspection of Fig.~\ref{Fig:Imaging}(a)-(c) clearly demonstrates that the design target for resolution with fixed magnification and geometrical constraints have been met as the contrast at the nominal ion-ion distance of $\approx$\SI{10}{\micro\meter} or 50~cycles per mm is close to the unaberrated case for all field points and remains high for magnified crystal sizes surpassing the available detector areas.

\begin{table}[H]%
\begin{center}
\renewcommand{\arraystretch}{1.25}

\begin{minipage}[t]{0.49\columnwidth}
\
\begin{center}

\begin{tabular}{c| c c c}
 & $x$ (mm) & $\theta$ (\textdegree) & $\phi$ (\textdegree) \\\hline
T1 & 2.22 & 90.0 & 90.0 \\
T2 & 7.30 & 121.5 & 90.0 \\
T3 & 28.10 & 90.0 & 105.4 \\
T4 & 27.56 & 105.4 & 90.0 \\
T5 & 10.73 & 90.0 & 90.0 \\\hline
T6 & 44.13 & 80.0 & 80.0
\end{tabular}
\end{center}
\end{minipage}
\begin{minipage}[t]{0.49\columnwidth}
\
\begin{center}

\begin{tabular}{c| c c c}
 & $x$ (mm) & $\theta$ (\textdegree) & $\phi$ (\textdegree) \\\hline
S1 & 3.85 & 104.3 & 90.0 \\
S2 & 4.25 &100.3 & 94.0 \\
S3 & 14.08 & 94.0 & 90.0 
\end{tabular}
\end{center}
\end{minipage}

\caption{Lens spacer parameters for topview and sideview imaging system from Section~\ref{Sec:Imaging} as shown in Fig.~\ref{Fig:Imaging}. Distance between lens contact points $x$ and lens contact angles $\phi$ and $\theta$. Spacer thickness $d$ for topview spacers is $d=\SI{1.5}{\milli\meter}$ and for sideview spacers $d=\SI{1.0}{\milli\meter}$. Spacer outer diameter is 2" for topview spacers and 1" for sideview spacers, except for T6 which is 1" and has $d=\SI{2.5}{\milli\meter}$. Note that spacer T1 has an additional chamfer to avoid collision with the meniscus lens. }
\label{Table:Spacers}
\end{center}
\end{table}


\end{widetext}

\end{document}